\documentclass[%
 preprint, 
showpacs,preprintnumbers,
 amsmath,amssymb,
 aps,color,nofootinbib
]{revtex4-1}
\usepackage[pdftex]{graphicx,color}
\graphicspath{{./Figures/}}
\usepackage{subfigure,
mathrsfs,braket,amsmath,amssymb,dcolumn,bm,comment}
\newlength{\subfigwidth}
\newlength{\subfigcolsep}
\setkeys{Gin}{width=\subfigwidth}
\makeatletter 
 
\@addtoreset{figure}{section} 
\makeatother 
\begin{document}
\preprint{}
\def\b{\begin{eqnarray}}
\def\e{\end{eqnarray}}
\def\aTr{{\rm Tr}}
\def\bpi{\bar{\pi}}
\def\tbr{\textcolor{red}}
\def\tcr{\textcolor{red}}
\def\ov{\overline}
\def\dprime{{\prime \prime}}
\def\nn{\nonumber}
\def\f{\frac}
\def\p{\partial}
\def\H{\mathcal{H}}
\def\VLL{\mathrm{VLL}}
\def\LR{\mathrm{LR}}
\def\SLL{\mathrm{SLL}}
\def\beq{\begin{equation}}
\def\eeq{\end{equation}}
\def\bea{\begin{eqnarray}}
\def\eea{\end{eqnarray}}
\def\bsub{\begin{subequations}}
\def\esub{\end{subequations}}
\def\dc{\stackrel{\leftrightarrow}{\partial}}
\def\d{\partial}
\def\sla#1{\rlap/#1}
\def\nn{\nonumber}
\def\bei{\begin{itemize}}
\def\eei{\end{itemize}}
\def\s{\partial \hspace{-.47em}/}
\def\ad{\overleftrightarrow{\partial}}
\def\para{%
\setlength{\unitlength}{1pt}%
\thinlines %
\begin{picture}(12, 12)%
\put(0,0){/}
\put(2,0){/}
\end{picture}%
}%
\title{Semi-aligned two Higgs doublet model}
\author{Naoyuki Haba}
\email[E-mail: ]{haba@riko.shimane-u.ac.jp}
\author{Hiroyuki Umeeda}
\email[E-mail: ]{umeeda@riko.shimane-u.ac.jp}
\author{Toshifumi Yamada}
\email[E-mail: ]{toshifumi@riko.shimane-u.ac.jp}
\affiliation{Graduate School of Science and Engineering, Shimane University, Matsue 690-8504, Japan}
\date{\today}
\begin{abstract}
In the left-right symmetric model based on $SU(2)_L\times SU(2)_R\times U(1)_{B-L}$ gauge symmetry,
 there appear heavy neutral scalar particles mediating quark flavor changing neutral currents (FCNCs) at tree level.
We consider a situation where such FCNCs give the only sign of the left-right model while $W_R$ gauge boson is decoupled,
and name it ``semi-aligned two Higgs doublet model" because the model resembles a two Higgs doublet model
 with mildly-aligned Yukawa couplings to quarks.
We predict a correlation among processes induced by quark FCNCs in the model,
 and argue that future precise calculation of meson-antimeson mixings and CP violation therein
 may hint at the semi-aligned two Higgs doublet model and the left-right model behind it.
\end{abstract}
\maketitle
\clearpage
\section{Introduction}
The left-right symmetric model~\cite{lr} based on $SU(2)_L\times SU(2)_R\times U(1)_{B-L}$ gauge symmetry and the left-right parity
 (symmetry under the Lorentzian parity transformation accompanied by the exchange of $SU(2)_L$ and $SU(2)_R$)
 is a well-motivated extension of the Standard Model (SM).
In the model, the chiral nature of the SM is beautifully attributed to spontaneous breaking of $SU(2)_R\times U(1)_{B-L}$ gauge symmetry.
More importantly, the model offers a new look into the strong CP problem;
 since the left-right parity demands the Yukawa couplings to be Hermitian and forbids gluon $\theta$ term at tree level,
 if one could find a symmetry-based reason that the two vacuum expectation values (VEVs) of the $SU(2)_L\times SU(2)_R$ bi-fundamental scalar are both made real,
 the strong CP problem would be solved.
Put another way, the original strong CP problem, which is about a miraculous cancellation between the $\theta$ term in QCD and the quark mass phases in the theory of electroweak symmetry breaking,
 is simplified to the issue of why the scalar potential does not break CP spontaneously.
Thus, the strong CP problem becomes more tractable, although the left-right model by itself does not solve it.

The first experimental hint of the left-right model would probably come in the form of quark flavor changing neutral currents (FCNCs) mediated by heavy neutral scalar particles at tree level,
 because the $SU(2)_L\times SU(2)_R$ bi-fundamental scalar necessarily has two unaligned Yukawa couplings both contributing to
 up and down-type quark masses to accommodate the Cabibbo-Kobayashi-Maskawa (CKM) matrix,
 and the resultant quark FCNCs and CP violation are efficiently
searched for through meson-antimeson mixings~\cite{lhc1,lhc2,old,zhang}.
On the other hand, recent studies have centered on the possibility of a direct measurement of $W_R$ gauge boson at the LHC,
 by elaborating a scalar potential where $SU(2)_R\times U(1)_{B-L}$ breaking VEV $v_R$ is below $\sim$5~TeV (hence $W_R$ mass being several TeV), while the heavy scalar particles have masses above $\sim$20~TeV to evade the bounds on FCNCs~\cite{lhc1,lhc2}.
Such a scalar potential contains $\mathcal{O}(1)$ quartic couplings that quickly become non-perturbative along renormalization group (RG) evolutions~\cite{chakrabortty,lrhiggs},
 and even if this is circumvented, there is no theoretical reason that favors having specially heavy scalar particles.
Conversely, if $W_R$ gauge boson has a mass similar to or larger than the heavy scalar particles
 (e.g. ($W_R$ mass)$\gtrsim$(heavy scalar mass)$=100$~TeV, so that the model evades the constraints from meson-antimeson mixings),
then $W_R$ gauge boson leaves no direct or indirect experimental signature and it is the heavy scalar particles that allow us to probe the left-right model through FCNCs they mediate
\footnote{
For a collider study of the heavy scalar particles in the left-right model, see Ref.\ \cite{Dev:2016dja}.
}.
\par
In this paper,
therefore, 
we pose the following question:
If quark FCNCs mediated by the heavy neutral scalars give the only sign of the left-right model, can we test the model?
To answer this, we extract the bi-fundamental scalar part of the model, assuming that $W_R$ gauge boson is decoupled from phenomenology,
 and systematically study its indirect signatures and their correlation.
We coin the term ``semi-aligned two Higgs doublet model (semi-aligned 2HDM)" to describe the bi-fundamental scalar part,
  in light of the fact that the bi-fundamental scalar contains two $SU(2)_L$ doublet scalars and their quark Yukawa couplings are mildly aligned
 reflecting the smallness of CKM mixing angles.

Our analysis starts by realizing that the flavor and CP-violating couplings of the heavy scalar particles are uniquely determined as follows:
Since the search for neutron electric dipole moment (EDM) has put a severe bound on the strong CP phase,
 we may concentrate on the limit with a vanishing spontaneous CP phase for the bi-fundamental scalar VEVs~\cite{strongcp}
\footnote{In this paper, we focus on phenomenological consequences of a vanishing spontaneous CP phase and do not discuss its theoretical origin.
For attempts to derive the vanishing spontaneous CP phase in the framework of the left-right model, see Ref.~\cite{kuchimanchi}.
}.
In this limit, the quark mass matrices are Hermitian and the mixing matrix for right-handed quarks is identical with the CKM matrix,
 which allows one to express the couplings of the heavy scalar particles to quarks in terms of the SM quark masses and CKM matrix,
 without free parameters.
The amplitudes for various FCNC processes are then calculated as functions of just one free parameter,
 that is, the nearly degenerate mass of the heavy scalar particles, and we thus predict a correlation among various FCNC processes.
We will show that indirect CP violation in kaon system, Re~$\epsilon$, the $B_d^0$ mass splitting, $\Delta M_{B_d}$, and 
 the $B_s^0$ mass splitting, $\Delta M_{B_s}$, are the most sensitive probes for the semi-aligned 2HDM,
 and if uncertainties in their calculation are reduced and 
 the prediction including the contributions of heavy scalar particles converges to the experimental values
 for some unique value of the heavy scalar mass,
 it is evidence for the semi-aligned 2HDM and the left-right model behind it.

A novelty of our study compared to previous works~\cite{lhc1,lhc2}
 is that we concentrate on the limit with decoupled $W_R$ gauge boson, which enables us to compute all amplitudes with only one free parameter
 and investigate their correlation.
Also, we pay attention to the fact that new physics contributions can distort the determination of the CKM matrix.
To avoid this, when we derive the current bound on the semi-aligned 2HDM, we refit the CKM matrix,
 attempting to fit the experimental data with SM+new physics contributions and using a tension in the fitting to constrain the model.
When we make a prediction for FCNC processes, we assume that the CKM matrix is determined beforehand in a way unaffected by new physics contributions.

This paper is organized as follows:
In Section~II, we describe the semi-aligned 2HDM induced from the left-right model,
 and calculate the mass and couplings of the heavy scalar particles.
In Section~III, we review the procedure for computing amplitudes for $\Delta F=2$ processes.
In Section~IV, we derive the current bound on the heavy neutral scalar mass.
Section~V presents our main results, which are Re~$\epsilon$, $\Delta M_{B_d}$ and $\Delta M_{B_s}$ expressed in terms of one parameter.
Section~VI summarizes the paper.
\\
\section{Model}

We start from the left-right symmetric model based on $SU(3)_C \times SU(2)_L\times SU(2)_R \times U(1)_{B-L}$ gauge symmetry~\cite{lr}
 and the left-right parity, whose full expression is in Appendix~A.
The VEV of the $SU(2)_R$ triplet scalar, $v_R$, breaks $SU(2)_R\times U(1)_{B-L}$ into hypercharge $U(1)_Y$ and also breaks the left-right parity.
In this paper, we focus on a limit where $v_R$ is much larger than the electroweak scale, and at the same time, the quartic couplings between two $SU(2)_R$ triplets and two bi-fundamentals 
 are much smaller than 1, namely,
\begin{align} 
v_R&\gg v\simeq246~{\rm GeV}, \ \ \ \ \ \vert\alpha_1\vert,~\vert\alpha_{2R}\vert,~\vert\alpha_{2I}\vert,~\vert\alpha_3\vert\ll1,
\label{limit}
 \end{align}
 where $\alpha_1,~\alpha_{2R},~\alpha_{2I},~\alpha_3$ are defined in the Lagrangian of Appendix~A.
In the limit with Eq.~(\ref{limit}), the low-energy theory at scales below $v_R$ (we call it the semi-aligned 2HDM)
 contains the SM fermions+three right-handed neutrinos+the bi-fundamental scalar and possesses SM $SU(3)_C \times SU(2)_L \times U(1)_Y$ gauge symmetry.
However, the Yukawa couplings and self-couplings of the bi-fundamental scalar respect 
 global $SU(2)_R$ symmetry and the left-right parity, which are remnants of the left-right model,
 and hence have highly constrained structures.
In Table~\ref{content}, we summarize the fields in the semi-aligned 2HDM
 and their charges in $SU(3)_C \times SU(2)_L \times U(1)_Y$ gauge group and global $SU(2)_R$ group.
\begin{table}[h]
\caption{Field content and charge assignments. $i=1,2,3$ is the flavor index.}
\begin{center}
\begin{tabular}{|c|c|c|c|c|c|c|} \hline
Field    & Lorentz $SO(1,3)$ & $SU(3)_C$ & $SU(2)_L$ & global $SU(2)_R$ & $U(1)_Y$ \\ \hline \hline
$q_L^i$& ({\bf 2},~{\bf 1})   &{\bf 3}          &{\bf 2}     &{\bf 1}                    &  1/6  \\ \hline
$q_R^i= \begin{pmatrix} 
      u_R^i \\
      d_R^i \\
   \end{pmatrix}$& ({\bf 1},~{\bf 2})  &{\bf 3}  &{\bf 1}        &{\bf 2}                 &     $\begin{pmatrix} 
         2/3 \\
         $-1/3$ \\
      \end{pmatrix}$  \\ \hline
$\ell_L^i$& ({\bf 2},~{\bf 1})   &{\bf 1}          &{\bf 2}        &{\bf 1}                 &  $-1/2$  \\ \hline
$\ell_R^i=\begin{pmatrix} 
      \nu_R^i \\
      e_R^i \\
   \end{pmatrix}$& ({\bf 1},~{\bf 2})               &{\bf 1}          &{\bf 1}                &{\bf 2}         &  $\begin{pmatrix} 
         0 \\
         $-1$ \\
      \end{pmatrix}$  \\ \hline
$\Phi=(i\sigma_2 H_u^*, \ i\sigma_2 H_d^*)$& {\bf 1}    &    ${\bf 1}$      & {\bf 2}          & {\bf 2}         &  $(-1/2, \ 1/2)$  \\ \hline
\end{tabular}
\end{center}
\label{content}
\end{table}
Here, the SM right-handed fermions form doublets of global $SU(2)_R$ symmetry, $q_R$ and $\ell_R$.
The bi-fundamental scalar $\Phi$ is expressed as a $2\times2$ matrix transforming under a $SU(2)_L\times SU(2)_R$ gauge transformation as 
\begin{align} 
\Phi&\to e^{i\tau^a\theta_L^a}\Phi e^{-i\tau^a\theta_R^a}, \ \ \ \ \ \theta_L^a, \, \theta_R^a:{\rm gauge \ parameters}, \ \ \ \tau^a\equiv \sigma^a/2,
\end{align}
 which is then decomposed into two $SU(2)_L$ doublet scalars with hypercharge $Y=\pm1/2$, $H_u$ and $H_d$,
 as $\Phi=(i\sigma_2 H_u^*, \ i\sigma_2 H_d^*)$.
The Lagrangian of the semi-aligned 2HDM is given by
\begin{align} 
-{\cal L}&=(Y_q)_{ij} \ \bar{q}_L^i \Phi q_R^j + (\tilde{Y}_q)_{ij} \ \bar{q}_L^i \tilde{\Phi} q_R^j
+(Y_\ell)_{ij}\ \bar{\ell}_L^i \Phi \ell_R^j + (\tilde{Y}_\ell)_{ij}\ \bar{\ell}_L^i \tilde{\Phi} \ell_R^j + {\rm H.c.}
\label{yukawa}\\
&+\frac{v_R}{\sqrt{2}}(Y_M)_{ij} \ N_R^{i\,T} \epsilon N_R^j + {\rm H.c.}
\label{majoranayukawa}\\
&+m_1^2 \ {\rm tr}\left[\Phi^\dagger\Phi\right]
+m_{2R}^2 \ {\rm tr}\left[\Phi^\dagger\tilde{\Phi}+\Phi\tilde{\Phi}^\dagger\right]
+m_{2I}^2 \ i \ {\rm tr}\left[\Phi^\dagger\tilde{\Phi}-\Phi\tilde{\Phi}^\dagger\right]
+m_3^2 \ {\rm tr}\left[\Phi^\dagger\Phi    \begin{pmatrix} 
      1 & 0 \\
      0 & 0 \\
   \end{pmatrix}\right]
\nonumber \\
&+\lambda_1 \ {\rm tr}\left[\Phi^\dagger\Phi\right]^2
+\lambda_2 \ \left({\rm tr}\left[\Phi^\dagger\tilde{\Phi}\right]^2+{\rm tr}\left[\Phi\tilde{\Phi}^\dagger\right]^2\right)
+\lambda_3 \ {\rm tr}\left[\Phi^\dagger\tilde{\Phi}\right]{\rm tr}\left[\Phi\tilde{\Phi}^\dagger\right]
\nonumber \\
&+\lambda_4 \ {\rm tr}\left[\Phi^\dagger\Phi\right]{\rm tr}\left[\Phi^\dagger\tilde{\Phi}+\Phi\tilde{\Phi}^\dagger\right],
\label{scalarpotential}
\\
&{\rm with} \ \tilde{\Phi}\equiv i\sigma_2 \, \Phi^* \, i\sigma_2,
\nonumber
\end{align}
 where $m_1^2$, $m_{2R}^2$, $m_{2I}^2$, $m_3^2$, $\lambda_1,\lambda_2,\lambda_3,\lambda_ 4$ are all real,
 and the Yukawa coupling matrices are Hermitian, 
 $Y_q^\dagger= Y_q$, $\tilde{Y}_q^\dagger= \tilde{Y}_q$, $Y_\ell^\dagger= Y_\ell$, $\tilde{Y}_\ell^\dagger= \tilde{Y}_\ell$.
Notice that $m_{2I}^2$ softly breaks the left-right parity, and $m_3^2$ softly breaks $SU(2)_R$ symmetry.
Both result from the spontaneous left-right symmetry breaking and are proportional to $v_R^2$.

$\Phi$ develops a VEV to break $SU(2)_L\times U(1)_Y$ symmetry.
Through a $SU(2)_L$ plus $\sigma_3$ part of $SU(2)_R$ symmetry transformation
 \footnote{
 This symmetry transformation generates a phase for $v_R$, but this can be negated by a $U(1)_{B-L}$ symmetry transformation.
 }, 
 the VEV is made into the following form, with one VEV having a CP phase $\alpha$:
\begin{align} 
\langle \Phi \rangle &=   \frac{1}{\sqrt{2}}\begin{pmatrix} 
      v\sin\beta & 0 \\
      0 & -e^{i\,\alpha}\,v\cos\beta \\
   \end{pmatrix}, \ \ \ \ \ v\simeq246~{\rm GeV}, \ \ \ \sin\beta >0, \ \cos\beta>0.
\end{align}
The mass matrices for up-type quarks, $M_u$, down-type quarks, $M_d$, charged leptons, $M_e$, are given by
\begin{align} 
M_u &= \frac{v}{\sqrt{2}}(Y_q \, \sin\beta + \tilde{Y}_q \, \cos\beta\,e^{-i\,\alpha}),
\label{upmass}\\
M_d &= -\frac{v}{\sqrt{2}}(Y_q \, \cos\beta\,e^{i\,\alpha} + \tilde{Y}_q \, \sin\beta),
\label{downmass}\\
M_e &= -\frac{v}{\sqrt{2}}(Y_\ell \, \cos\beta\,e^{i\,\alpha} + \tilde{Y}_\ell \, \sin\beta).
\end{align}

The spontaneous CP phase $\alpha$ has already been severely constrained by the search for neutron EDM.
Since $W_R$ is decoupled and the coupling of scalar particles to up and down quarks is Yukawa-suppressed,
 perturbative corrections to neutron EDM are negligible,
 and the experimental bound is directly translated into a bound on $\arg\det(M_uM_d)$ and hence on $\alpha$.
In the limit of neglecting the quark flavor mixing (but no assumptions are made on $\beta$ or $\alpha$), we obtain, from Eqs.~(\ref{upmass}, \ref{downmass}), the following formula:
\begin{align} 
\arg\det(M_uM_d)&\simeq\theta_{ud}+\theta_{cs}+\theta_{tb},
\nonumber \\
\sin\theta_{ud}&=-\frac{m_u^2-m_d^2}{2m_um_d}\tan(2\beta)\sin\alpha,\ \ \ (u,d)\to(c,s), \ \ \ (u,d)\to(t,b).
\end{align}
The current experimental bound~\cite{nedm} roughly gives
\begin{align} 
10^{-10}&>\vert\bar{\theta}\vert=\vert\arg\det\left(M_uM_d\right)\vert\simeq\vert\theta_{ud}+\theta_{cs}+\theta_{tb}\vert,
\end{align}
 where it should be reminded that the QCD $\theta$ term is prohibited at tree level by the left-right parity.
$\alpha$ is thus constrained to be much below 1, and based on this fact, we fix $\alpha=0$ in the rest of the paper
\footnote{
Deriving $\alpha=0$ theoretically is equivalent to solving the strong CP problem,
 which we do not attempt in this paper.
}.

The physical scalar particles after the electroweak symmetry breaking are
\footnote{
Since $\alpha=0$, CP is not broken spontaneously and the CP-even and odd scalar particles can be defined.
}
 a charged scalar, $H^\pm$, a CP-odd scalar, $A$,
 a lighter CP-even scalar which we identify with the SM Higgs particle, $h$, and a heavier CP-even scalar, $H$.
The $H^\pm$ and $A$ masses read
\begin{align} 
m_{H^\pm}^2 &= \frac{m_3^2}{\cos2\beta},
\label{chmass} 
\\
m_A^2 &= \frac{m_3^2}{\cos2\beta}-(4\lambda_2-2\lambda_3)v^2,
\label{amass}
\end{align}
The $H$ and $h$ masses expanded to the order of $\mathcal{O}(v^2/\vert m_3^2\vert)$ are found to be
\begin{align} 
m_H^2 &\simeq \frac{m_3^2}{\cos2\beta}+(4\lambda_2+2\lambda_3)\cos^22\beta \ v^2,
\label{largehmass}
\\
m_h^2 &\simeq 2\lambda_1 v^2+(4\lambda_2+2\lambda_3)\sin^22\beta \ v^2 +4\lambda_4 \sin2\beta \ v^2.
\label{smallhmass}
\end{align}
The bi-fundamental scalar $\Phi=(i\sigma_2 H_u^*, \ i\sigma_2 H_d^*)$
 can be decomposed into the physical scalar particles $H^\pm,A,H,h$ and Nambu-Goldstone bosons, $G^\pm,G^0$, in the following way:
\begin{align} 
H_u &=   \begin{pmatrix} 
      -\sin\beta \ G^+ +\cos\beta \ H^+ \\
      \frac{1}{\sqrt{2}}\left(\sin\beta \ v+\cos\gamma \ h+\sin\gamma \ H -i \sin\beta \ G^0 +i\cos\beta \ A\right) \\
   \end{pmatrix},
   \nonumber \\
H_d &=   \begin{pmatrix} 
      \frac{1}{\sqrt{2}}\left(\cos\beta \ v-\sin\gamma \ h+\cos\gamma \ H +i \cos\beta \ G^0 +i\sin\beta \ A\right) \\
       \cos\beta \ G^- +\sin\beta \ H^- \\
   \end{pmatrix},
\end{align}
 where $\gamma$ is the mixing angle of the CP-even scalars satisfying
\begin{align} 
\tan2\gamma &= \frac{m_3^2-2v^2(\lambda_1+2\lambda_2+\lambda_3)\cos2\beta-4v^2 \lambda_4 \cot2\beta}
{m_3^2-2v^2(\lambda_1-2\lambda_2-\lambda_3)\cos2\beta}\tan2\beta, \ \ \ \ \ \ 0>\gamma >-\pi/2.
\end{align}
Since the $A$ and $H$ masses are experimentally constrained to be above $\sim$10~TeV,
 we work in the decoupling limit with $v^2/\vert m_3^2\vert \to 0$ to realize $m_H^2, m_A^2 \gg m_h^2$,
 in which case the masses of $H^\pm,A,H$ and the CP-even scalar mixing angle $\gamma$ satisfy
\begin{align} 
m_{H^\pm}^2&=m_A^2=m_H^2,
\label{massequality}
\\
\gamma &= \beta-\frac{\pi}{2}.
\label{alphabeta}
\end{align}

Equation~(\ref{yukawa}) induces Yukawa couplings for quarks and $H^\pm$, $H$ and $A$, given by
\begin{align} 
-{\cal L}_{yukawa} &= 
(Y_q\sin\beta-\tilde{Y}_q\cos\beta)_{ij} \, \bar{u}_L^i d_R^j \ H^+
+(-Y_q\cos\beta+\tilde{Y}_q\sin\beta)_{ij} \, \bar{d}_L^i u_R^j \ H^-
+{\rm H.c.}
\nonumber \\
&+\frac{i}{\sqrt{2}}(-Y_q\cos\beta+\tilde{Y}_q\sin\beta)_{ij} \ \bar{u}^i \gamma_5 u^j \, A
+\frac{i}{\sqrt{2}}(Y_q\sin\beta-\tilde{Y}_q\cos\beta)_{ij} \ \bar{d}^i \gamma_5 d^j \, A
\nonumber \\
&+\frac{1}{\sqrt{2}}(Y_q\sin\gamma+\tilde{Y}_q\cos\gamma)_{ij} \ \bar{u}^i u^j \, H
+\frac{1}{\sqrt{2}}(-Y_q\cos\gamma-\tilde{Y}_q\sin\gamma)_{ij} \ \bar{d}^i d^j \, H
\\
&\simeq -\sqrt{2}\frac{(M_u+M_d \sin2\beta)_{ij}}{v \, \cos2\beta}\bar{u}_L^i d_R^j H^+
+\sqrt{2}\frac{(M_d+M_u \sin2\beta)_{ij}}{v \, \cos2\beta}\bar{d}_L^i u_R^j H^-
+{\rm H.c.}
\nonumber \\
&+\frac{(M_d+M_u \sin2\beta)_{ij}}{v \, \cos2\beta}\bar{u}^i u^j H
+i\frac{(M_d+M_u \sin2\beta)_{ij}}{v \, \cos2\beta}\bar{u}^i \gamma_5 u^j A
\nonumber \\
&+\frac{(M_u+M_d \sin2\beta)_{ij}}{v \, \cos2\beta}\bar{d}^i d^j H
-i\frac{(M_u+M_d \sin2\beta)_{ij}}{v \, \cos2\beta}\bar{d}^i \gamma_5 d^j A,
\label{yukawaint}
\end{align}
 where Eq.~(\ref{alphabeta}) has been used.
We diagonalize the quark mass matrices by the rotation,
\begin{align} 
&u_{L, R}\to V_u u_{L, R}, \ \ \ \ \ d_{L, R}\to V_d d_{L, R},
\label{trans}
\end{align}
 to obtain
\begin{align}
&V^\dagger_u M_u V_u=m_u^D, \ \ \ \ \ V_d^\dagger M_d V_d=m_d^D,
\label{diageq}\\
&m_u^D=\mathrm{diag}(s_u m_u, s_c m_c, s_t m_t), \ \ \ \ \ m_d^D=\mathrm{diag}(s_d m_d, s_s m_s, s_b m_b),
\label{Eq:mass}
\end{align}
 where $m_u^D$ and $m_d^D$ are diagonalized mass matrices for up and down-type quarks, respectively, 
 and $s_f=\pm 1 (f=u, c, t, d, s, b)$, which reflects sign uncertainty of the mass eigenvalues.
Note that the left and right-handed quarks are rotated with the same unitary matrix, since $\alpha=0$ and the mass matrices are Hermitian.
Accordingly, the Yukawa couplings become
\begin{align} 
-{\cal L}_{yukawa} &=
-\sqrt{2}\frac{(m_u^D V+V^\dagger m_d^D \sin2\beta)_{ij}}{v \, \cos2\beta}\bar{u}_L^i d_R^j H^+
+\sqrt{2}\frac{(m_d^D V^\dagger +V m_u^D \sin2\beta)_{ij}}{v \, \cos2\beta}\bar{d}_L^i u_R^j H^-
+{\rm H.c.}
\nonumber\\
&+\frac{(V m_d^D V^\dagger+m_u^D \sin2\beta)_{ij}}{v \, \cos2\beta}\bar{u}^i u^j H
+i\frac{(V m_d^D V^\dagger+m_u^D \sin2\beta)_{ij}}{v \, \cos2\beta}\bar{u}^i \gamma_5 u^j A
\nonumber \\
&+\frac{(V^\dagger m_u^D V+m_d^D \sin2\beta)_{ij}}{v \, \cos2\beta}\bar{d}^i d^j H
-i\frac{(V^\dagger m_u^D V+m_d^D \sin2\beta)_{ij}}{v \, \cos2\beta}\bar{d}^i \gamma_5 d^j A,
\label{yukawafinal}
\end{align}
 where we have defined the CKM matrix, $V$, as
\begin{align}
V&=V_u^\dagger V_d.
\end{align}
We find that the following part in Eq.~(\ref{yukawafinal}) induces
 FCNCs at tree level by the exchange of heavy neutral scalars $H,A$, with the strength controlled by the CKM matrix multiplied by quark masses:
\begin{align} 
-{\cal L}_{yukawa} &\supset
\frac{(V m_d^D V^\dagger)_{ij}}{v \, \cos2\beta}\left(\bar{u}^i u^j H+i\bar{u}^i \gamma_5 u^j A\right)
+\frac{(V^\dagger m_u^D V)_{ij}}{v \, \cos2\beta}\left(\bar{d}^i d^j H-i\bar{d}^i \gamma_5 d^j A\right).
\label{FCNC}
\end{align}
Flavor violation is suppressed by off-diagonal components of the CKM matrix, and possibly by light quark masses,
 which is a characteristic property of the model.

Since the Yukawa couplings always appear in combination with the factor $1/\cos2\beta$, hereafter we redefine the heavy scalar masses as
\footnote{
If we require $\tan\beta\sim m_t/m_b$ so that the top and bottom quark mass ratio is derived without fine-tuning,
 we have $\cos^22\beta\simeq1$ and this redefinition becomes trivial.
}
\begin{align} 
{\rm new} \ m_{H^{\pm}}^2&\equiv m_{H^{\pm}}^2 \cos^22\beta, \ \ \
{\rm new} \ m_H^2\equiv m_H^2 \cos^22\beta, \ \ \
{\rm new} \ m_A^2\equiv m_A^2 \cos^22\beta.
\end{align}

As a reference, we present the absolute values of the flavor-violating part of the Yukawa couplings
 for $s_f=+1 (f=u, c, t, d, s, b)$:
\begin{align}
\mathrm{For\ up}-\mathrm{type}:\quad
\displaystyle\frac{|V m_d^DV^\dagger|}{v}&=
\begin{pmatrix}
 0.000037& 0.000080 & 0.000059 \\
 * & 0.00038 & 0.00068 \\
 * & * & 0.017 \\
\end{pmatrix},\label{UPFCNC}\\
\mathrm{For\ down}-\mathrm{type}:\quad
\displaystyle\frac{|V^\dagger m_u^DV|}{v}&=
\begin{pmatrix}
 0.00032 & 0.0013& 0.0060 \\
 * & 0.0058 & 0.027 \\
 * & * & 0.70 \\
\end{pmatrix}.
\label{DOWNFCNC}
\end{align}
The off-diagonal components in the above matrices indicate the strength of FCNCs mediated by the neutral scalars.
Here, the CKM matrix components are obtained from the Wolfenstein parameters reported by the CKMfitter, which read~\cite{Charles:2015gya},
\bea
\lambda=0.22548,\qquad A=0.810,\qquad
\bar{\rho}=0.145,\qquad \bar{\eta}=0.343.\label{Eq:CKMf}
\eea
\\

We comment in passing that the Yukawa couplings for leptons and $H^\pm$, $H$ and $A$
 are obtained by a simple replacement: $d\to e,\, u\to \nu,\, M_d\to M_e,\, M_u\to M_D$ in Eq.~(\ref{yukawaint}), with $M_D$
 being the neutrino Dirac mass involved in the seesaw mechanism.
Due to our ignorance of the seesaw scale, we cannot predict the strength of the flavor-violating couplings for charged leptons.
\\

\section{$\Delta F=2$ amplitudes}
\begin{figure}[h!]
\includegraphics[width=5.5cm]{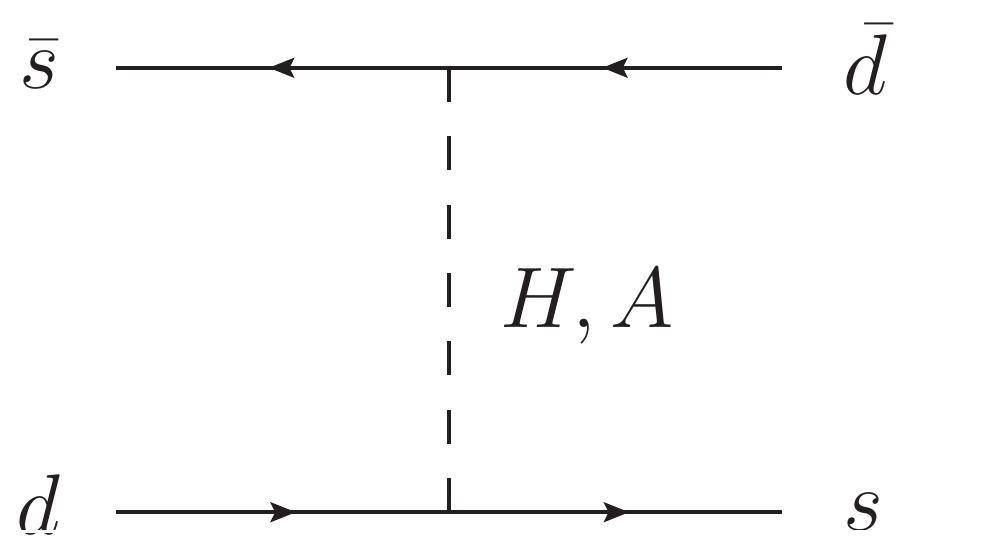}
\caption{
Diagram for $\Delta S=2$ process arising from the exchange of a heavy neutral scalar particle $H,A$,
  which we call ``flavor changing neutral Higgs (FCNH)".
Another crossed diagram is omitted.}
\label{Fig:1}
\end{figure}

We give formulae
for $\Delta F=2$ amplitudes to analyze
flavor observables in the quark sector.
In particular, mass differences for $K, B_d^0$ and $B_s^0$,
and a CP violating observable in kaon system are given.
The effective Hamiltonian contributing to $\Delta S=2$
processes is given as follows,
\bea
&\mathcal{H}_{\Delta S=2}^{\mathrm{FCNH}}
=\displaystyle\frac{G_F}{\sqrt{2}}
\left(C_{S} O_{S}+C_{P} O_P\right)+\mathrm{H.c.},&
\label{ham}\\
&C_S=-\displaystyle\frac{1}{m_H^2}\left[\displaystyle\sum_{k}^{u, c, t} \lambda_k^{sd} (m_u^D)^k
\right]^2,\quad
C_P=\displaystyle\frac{1}{m_A^2}\left[\displaystyle\sum_{k}^{u, c, t} \lambda_k^{sd} (m_u^D)^k
\right]^2,
\quad
\lambda_k^{ij}=V_{ki}^* V_{kj},&\\
&O_S=\bar{s}d \bar{s}d,\qquad
O_P=\bar{s}\gamma_5d \bar{s}\gamma_5d.&\label{Heff}
\eea
Proper replacement of the indices in
Eqs.\ (\ref{ham}-\ref{Heff})
enables us to write the effective Hamiltonian in $\Delta B=2$ processes.
The Hamiltonian in Eq.\ (\ref{ham})
represents the contribution of
a tree-level diagram in Figure\ \ref{Fig:1} arising
from the exchange of a heavy neutral scalar particle $H,A$,
 which we denote ``flavor changing neutral Higgs (FCNH)".
When the heavy scalar particles are degenerate,
one can obtain a further simplified Hamiltonian.
Including the SM contribution, we can write,
\bea
&\mathcal{H}_{\Delta S=2}=\mathcal{H}_{\Delta S=2}
^{\mathrm{SM}}+\mathcal{H}_{\Delta S=2}
^{\mathrm{FCNH}},&\\
&\mathcal{H}_{\Delta S=2}^{\mathrm{SM}}=
C_1^{\mathrm{VLL}}Q_1^{\mathrm{VLL}}+\mathrm{H.c.},&
\label{Eq:effSM}\\
&\mathcal{H}_{\Delta S=2}^{\mathrm{FCNH}}=
\displaystyle\sum_{i=1}^{2}C_i^{\mathrm{LR}}Q_i
^{\mathrm{LR}}+\mathrm{H.c.,}&\label{Eq:eff}
\eea
where we follow the notation of Ref.\ \cite{Buras:2001ra}.
In Eq.\ (\ref{Eq:eff}), we do not include operators,
$Q_1^{\SLL}$, $Q_2^{\SLL}$ and other chirality-flipped ones
given in Ref.\ \cite{Buras:2001ra},
since they do not arise from the FCNH diagram.
Furthermore, $Q_1^{\SLL}$ and $Q_2^{\SLL}$
are decoupled from the mixing
with the other operators so that
we omit these contributions.
The Wilson coefficients and the operators
in Eqs.\ (\ref{Eq:effSM}, \ref{Eq:eff}) are given as,
\bea
&C_{1}^{\VLL}=
\displaystyle\frac{G_F^2 M_W^2}{(4\pi)^2}
4\tilde{S},\quad
C_{1}^{\LR}(\mu_H)=0,\quad
C_{2}^{\LR}(\mu_H)=
-\displaystyle\frac{2\sqrt{2}G_F}{m_H^2}
\left[\displaystyle\sum_{k}^{u, c, t}
\lambda_k^{sd} (m_u^D)^k\right]^2,\qquad&\label{Cs}\\
&Q_{1}^{\VLL}=
\bar{s}^\alpha\gamma^\mu P_Ld^\alpha
\bar{s}^\beta\gamma_\mu P_L d^\beta,
\quad
Q_{1}^{\LR}=\bar{s}^\alpha\gamma_\mu P_L d^\alpha
\bar{s}^\beta \gamma^\mu P_R d^\beta,\quad
Q_{2}^{\LR}=\bar{s}^\alpha P_L d^\alpha
\bar{s}^\beta P_R d^\beta,\qquad&\label{Op}
\eea
where $P_{R(L)}=(1\pm\gamma_5)/2$ denotes
chirality projection operators
while $\alpha$ and $\beta$ represent color indices.
In Eq.\ (\ref{Cs}),
$C_1^{\VLL}$ stands for the contribution within the SM,
and the Inami-Lim function \cite{Inami:1980fz} is given as,
\bea
\tilde{S}&=&\eta_1(\lambda^{sd}_c)^2S(x_c)+
\eta_2(\lambda^{sd}_t)^2S(x_t)+
2\eta_3\lambda^{sd}_c\lambda^{sd}_tS(x_c, x_t)
,\quad
x_i=\frac{m_i^2}{M_W^2}\;\; (i=c, t),\label{EqINAMI}\qquad\\
S(x_i, x_j)&=&x_ix_j\left[\frac{1}{x_i-x_j}\left(
\left(\frac{1}{4}
-\frac{3}{2}\frac{1}{x_i-1}
-\frac{3}{4}\frac{1}{(x_i-1)^2}\right)\ln x_i\right.\right.\nn\\
&&\quad\qquad\qquad
\;\: \left.\left.-\left(\frac{1}{4}
-\frac{3}{2}\frac{1}{x_j-1}
-\frac{3}{4}\frac{1}{(x_j-1)^2}\right)\ln x_j
\right)
-\frac{3}{4}\frac{1}{(x_i-1)(x_j-1)}
\right],\qquad
\\
S(x_i)&=&
\frac{4x_i-11x_i^2+x_i^3}{4(x_i-1)^2}+\frac{3}{2}
\left(\frac{x_i}{x_i-1}\right)^3\ln x_i,
\eea
where NLO QCD correction factors within the SM,
$(\eta_1, \eta_2, \eta_3)$,
have been calculated in Ref.\ \cite{Herrlich:1993yv}.
To be precise, one should multiply Eq.\ (\ref{EqINAMI})
by an overall factor which accounts
renormalization scale of lattice QCD calculation.\par
As for the $\Delta B=2$ processes,
we only take account of the contribution of internal top quarks,
and the corresponding NLO QCD correction is obtained
through the method in Ref.\ \cite{Buras:2001ra}.
The formulae for an anomalous dimension matrix
including two-loop contribution
are given in Ref.\ \cite{Buras:2000if}.
As remarked in the literature \cite{Buras:2001ra},
this renormalization group effect drastically enhances $C_{2}^{\LR}$
while it does not significantly change $C_1^{\LR}$.
In our analysis, new world averages of the QCD scale
obtained by PDG \cite{Patrignani:2016xqp} are used.
\par The matrix elements of the $\Delta F=2$
transition are parametrized as,
\bea
\bra{\bar{K^0}} Q_1^{\VLL}(\mu)\ket{K^0}
&=&\frac{1}{3}M_{K} f_{K}^2 B_1^{\VLL}(\mu),\label{Eq:mat1}\\
\bra{\bar{K^0}} Q_1^{\LR}(\mu)\ket{K^0}
&=&-\frac{1}{6}M_K f_K^2 B_1^{\LR}(\mu)
\left(\frac{M_K}{m_s(\mu)+m_d(\mu)}\right)^2,\\
\bra{\bar{K^0}} Q_2^{\LR}(\mu)\ket{K^0}
&=&\frac{1}{4}M_K f_K^2 B_2^{\LR}(\mu)
\left(\frac{M_K}{m_s(\mu)+m_d(\mu)}\right)^2,\\
\bra{\bar{B^0_q}} Q_1^{\VLL}(\mu)\ket{B^0_q}
&=&\frac{1}{3}M_{B_q} f_{B_q}^2 B_1^{q \VLL}(\mu),\\
\bra{\bar{B^0_q}} Q_1^{\LR}(\mu)\ket{B^0_q}
&=&-\frac{1}{6}M_{B_q} f_{B_q}^2 B_1^{q \LR}(\mu)
\left(\frac{M_{B_q}}{m_b(\mu)+m_q(\mu)}\right)^2,\\
\bra{\bar{B^0_q}} Q_2^{\LR}(\mu)\ket{B^0_q}
&=&\frac{1}{4}M_{B_q} f_{B_q}^2 B_2^{q \LR}(\mu)
\left(\frac{M_{B_q}}{m_b(\mu)+m_q(\mu)}\right)^2.\label{Eq:mat6}
\eea
where $q=d, s$.
In this normalization, kaon decay constant
is given as $f_K=156.1\ \mathrm{MeV}$.
The matrix elements in Eqs.\ (\ref{Eq:mat1}-\ref{Eq:mat6})
are written in terms of bag parameters, which represent the deviation from vacuum saturation approximation.
For these parameter,
we use the data which are calculated by the ETM collaboration
\cite{Carrasco:2013zta, Carrasco:2015pra}.
Their results are obtained in
$\overline{\mathrm{MS}}$ scheme,
and extracted from the result in the supersymmetric basis.
The correspondence
between bag parameters in the operator basis in Eq.\ (\ref{Op})
and ones in the supersymmetric basis
is given in Ref.\ \cite{Buras:2001ra}.
\par
The mass differences of neutral meson system
are obtained as follows,
\bea
&\Delta M_K= 2\mathrm{Re} M_{12},\quad
\Delta M_{B_q}= 2|M_{12}^q|,&\label{Eq:MassDif}\\
&M_{12}=M_{12}^{\mathrm{SM}}+M_{12}^{\mathrm{FCNH}},\quad
M_{12}^{q}=M_{12}^{q\mathrm{SM}}+M_{12}^{q\mathrm{FCNH}},&
\eea
where $M_{12}^{(q)}$ is divided into the SM part and the new physics part,
\bea
&M_{12}^{\mathrm{SM}}=\bra{\bar{K^0}}\mathcal{H}_{\Delta S=2}
^{\mathrm{SM}}\ket{K^0}^*,\quad
M_{12}^{\mathrm{FCNH}}=\bra{\bar{K^0}}\mathcal{H}_{\Delta S=2}
^{\mathrm{FCNH}}\ket{K^0}^*,&\label{M12K}\\
&M_{12}^{q\mathrm{SM}}=\bra{\bar{B^0_q}}\mathcal{H}_{\Delta B=2}
^{\mathrm{SM}}\ket{B^0_q}^*,\quad
M_{12}^{q\mathrm{FCNH}}=\bra{\bar{B^0_q}}\mathcal{H}_{\Delta B=2}
^{\mathrm{FCNH}}\ket{B^0_q}^*.&\label{M12}
\eea
Moreover, indirect CP violation
in kaon system is characterized by
\bea
\epsilon=\displaystyle\frac{e^{\frac{i\pi}{4}}}{\sqrt{2}}
\displaystyle\frac{\mathrm{Im}M_{12}}{\Delta M_K}.\label{Eq:epsilonK}
\eea
\section{Current bound on the model}
In this section, we obtain the bound on mass of heavy Higgs
through flavor observables.
First, $\sin2\beta_{\mathrm{eff}}$,
which represents CP violation of interference in
$B^0_d-\bar{B^0_d}$ mixing and $B^0_d\to J/\psi
K_S$, is analyzed.
As discussed later, the bounds
on Higgs mass are determined by p-values
of CKM fitting.
\par
In Table\ \ref{Tab:1}, input data which appear
in the numerical analysis are summarized.
\begin{table}[h!]
\caption{Input data used in the analysis.
The third column corresponds to
references on which the data are based.
For $K_L-K_S$ mass difference,
the fitting result w/ CPT assumption is given.
$\mathrm{Re}\ \epsilon$ is extracted from
asymmetry of semi-leptonic decay rates given by PDG.
}
\label{Tab:1}
\begin{tabular}{|c|c|c|}\hline
$\eta_1$ & $1.32^{+0.21}_{-0.23}$&\cite{Herrlich:1993yv}\\\hline
$\eta_2$ & $0.57^{+0.00}_{-0.01}$&\cite{Herrlich:1993yv}\\\hline
$\eta_3$ & $0.47^{+0.03}_{-0.04}$&\cite{Herrlich:1993yv}\\\hline
$\Lambda_{\overline{MS}}^{(6)}$ & $(87\pm 7)\ \mathrm{MeV}$
&\cite{Patrignani:2016xqp}\\\hline
$\Lambda_{\overline{MS}}^{(5)}$ & $(210\pm 15)\ \mathrm{MeV}$
&\cite{Patrignani:2016xqp}\\\hline
$\Lambda_{\overline{MS}}^{(4)}$ & $(291\pm 19)\ \mathrm{MeV}$
&\cite{Patrignani:2016xqp}\\\hline
$\mathrm{Re}\ \epsilon$ & $(1.66\pm 0.03)\times10^{-3}$
&\cite{Patrignani:2016xqp}\\\hline
$\Delta M_K$ & $3.484 \pm 0.006\  [10^{-12}\ \mathrm{MeV}]$
&\cite{Patrignani:2016xqp}\\\hline
$\Delta M_{B_d}$ & $0.5064\pm 0.0019\ [\mathrm{ps}^{-1}]$
&\cite{Amhis:2016xyh}\\\hline
$\Delta M_{B_s}$ & $17.757\pm0.0021\ [\mathrm{ps}^{-1}]$
&\cite{Amhis:2016xyh}\\\hline
$\sin 2\beta_{\mathrm{eff}}$ & $0.691\pm 0.017$
&\cite{Amhis:2016xyh}\\\hline
$B_1^{\VLL} (3\ \mathrm{GeV})$
& $0.506\pm0.017$&\cite{Carrasco:2015pra}
\\\hline
$B_1^{\LR} (3\ \mathrm{GeV})$
& $0.49\pm0.04$&\cite{Carrasco:2015pra}
\\\hline
$B_2^{\LR} (3\ \mathrm{GeV})$
& $0.78\pm0.05$&\cite{Carrasco:2015pra}
\\\hline
$f_{B_d}\sqrt{B_1^{d\VLL}}(m_b)$
& $174\pm8\ \mathrm{MeV}$&\cite{Carrasco:2013zta}
\\\hline
$f_{B_d}\sqrt{B_1^{d\LR}}(m_b)$
& $229\pm14\ \mathrm{MeV}$&\cite{Carrasco:2013zta}
\\\hline
$f_{B_d}\sqrt{B_2^{d\LR}}(m_b)$&
$185\pm9\ \mathrm{MeV}$&\cite{Carrasco:2013zta}
\\\hline
$f_{B_s}\sqrt{B_1^{s\VLL}}(m_b)$
& $211\pm8\ \mathrm{MeV}$&\cite{Carrasco:2013zta}
\\\hline
$f_{B_s}\sqrt{B_1^{s\LR}}(m_b)$
& $285\pm14\ \mathrm{MeV}$&\cite{Carrasco:2013zta}
\\\hline
$f_{B_s}\sqrt{B_2^{s\LR}}(m_b)$&
$220\pm9\ \mathrm{MeV}$&\cite{Carrasco:2013zta}
\\\hline
\end{tabular}
\end{table}
\subsection{$\sin2\beta_{\mathrm{eff}}$ measured in
$B^0_d\to
J/\psi K_S$ decay}
Throughout this paper, we assume that direct CP violation
in $B^0_d\to J/\psi K_S$ decay is negligible.
Within this approximation,
the time-dependent CP asymmetry in $B^0_d\to
J/\psi K_S$ decay is given as follows,
\bea
A_{CP}(t)&\simeq& S_{J/\psi K_S} \sin (\Delta M_{B_d} t),\\
S_{J/\psi K_S}&\simeq&\mathrm{Im}\left(\frac{q}{p}\frac{\bar{A}_{J/\psi K_S}}{A_{J/\psi K_S}}\right),\label{Eq:SpsiKs}
\eea
where $q/p$ is a mixing parameter \cite{REVIEW}
in $B^0_d$ system
while $A_{J/\psi K_S} (\bar{A}_{J/\psi K_S})$
represents decay amplitude of $B_d^0 (\bar{B_d^0})\to J/\psi K_S$.
In the semi-aligned 2HDM,
the decay amplitude in Eq.\ (\ref{Eq:SpsiKs}) does not
deviate from the SM prediction, since a diagram of the charged scalar exchange
gives rise to minor modification due to smallness of Yukawa couplings.
We do not take account of such negligible contribution
for the decay amplitude.
Furthermore, the correction coming from penguin
pollution in the SM is also small \cite{Li:2006vq}
because of the suppression for the CKM and the loop factor, and hence we ignore this effect.
Meanwhile, the mixing parameter for $B_d^0$ system,
$q/p$, and one for kaon system are modified
due to the diagrams for FCNH exchange.
Hence, the parameter in Eq.\ (\ref{Eq:SpsiKs})
is given as \cite{Ball:1999mb, Kiers:2002cz},
\bea
\sin2\beta_{\mathrm{eff}}=S_{J/\psi K_S}&=&\sin\left[2\beta+
\mathrm{arg}\left(1+\frac{M_{12}^{d \mathrm{FCNH}}}{M_{12}
^{d \mathrm{SM}}}\right)
-\mathrm{arg}\left(1+\frac{M_{12}^{\mathrm{FCNH}}}{M_{12}
^{\mathrm{SM}}}\right)\right],\\
\beta&=&\mathrm{arg}
\left(-\frac{V_{cd}V_{cb}^*}{V_{td}V_{tb}^*}
\right),\label{Eq:beta}\eea
where the definitions of $M_{12}^{(d)\mathrm{SM}}$
and $M_{12}^{(d)\mathrm{FCNC}}$ are given in Eqs.\ (\ref{M12K},
\ref{M12}). Thus, the experimental observable
deviates from $\sin2\beta$ due to modification of the mixing parameters.
\begin{figure}[h!]
\includegraphics[width=8.2cm]{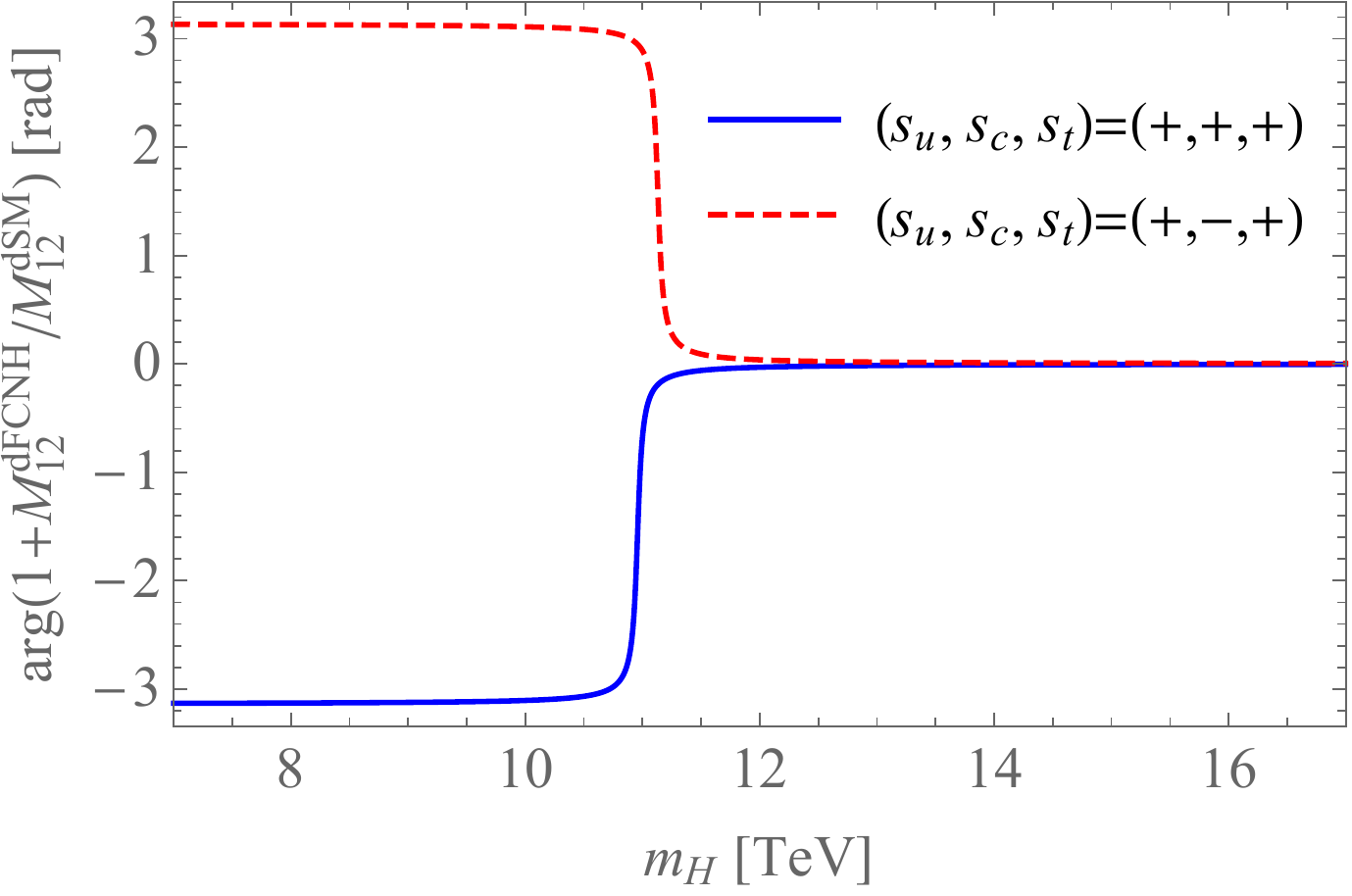}
   \label{Data9}
\caption{Argument of $1+M_{12}^{d \mathrm{FCNH}}/
M_{12}^{d \mathrm{SM}}$.}
\label{Fig:4}
\end{figure}
\begin{figure}[h!]
  \setlength{\subfigwidth}{.5\linewidth}
  \addtolength{\subfigwidth}{-.5\subfigcolsep}
  \begin{minipage}[b]{\subfigwidth}
    \subfigure[]{\includegraphics[width=8.2cm]{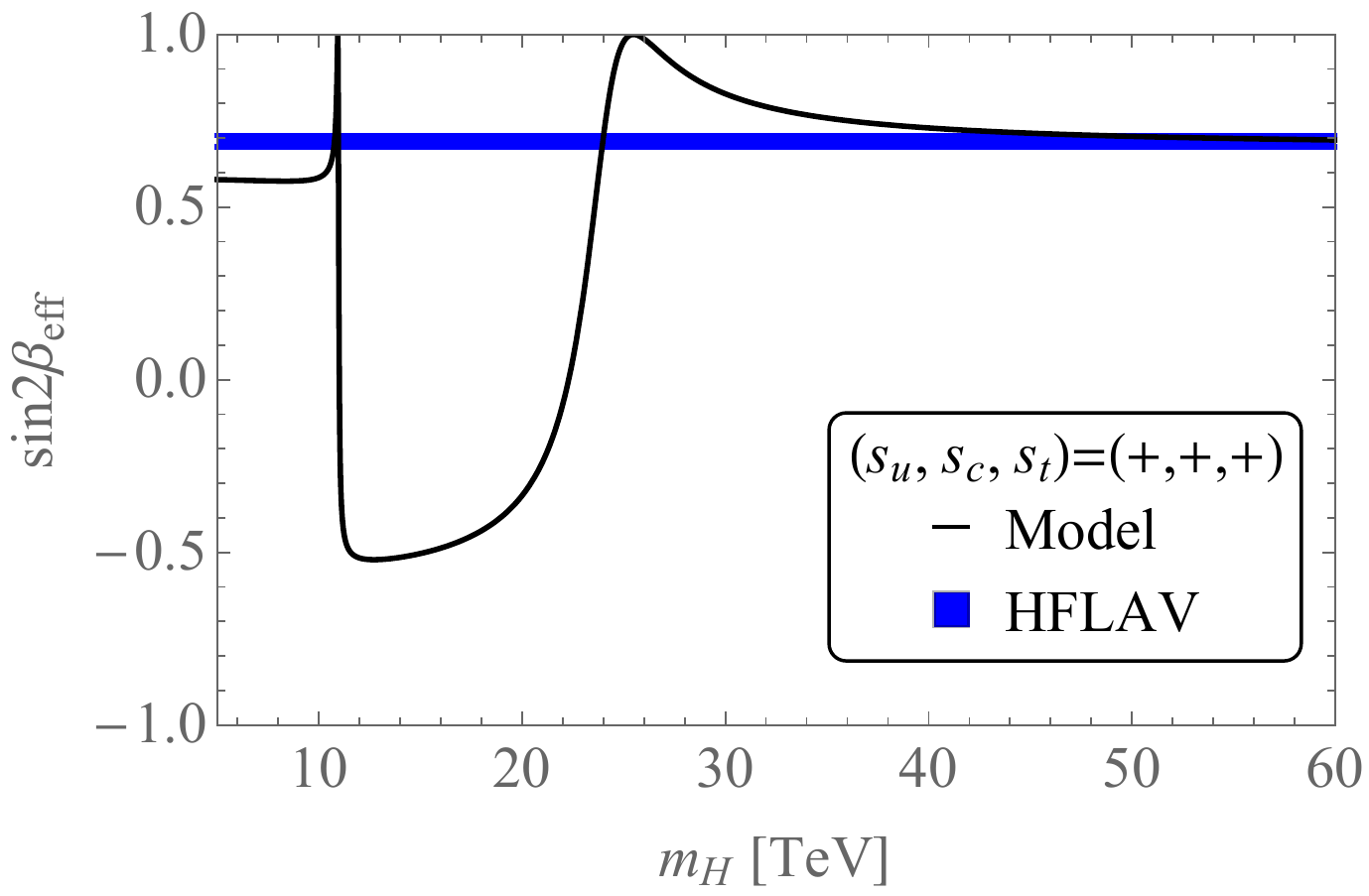}
   \label{Data9}}
  \end{minipage}
  \begin{minipage}[b]{\subfigwidth}
    \subfigure[]{\includegraphics[width=8.2cm]{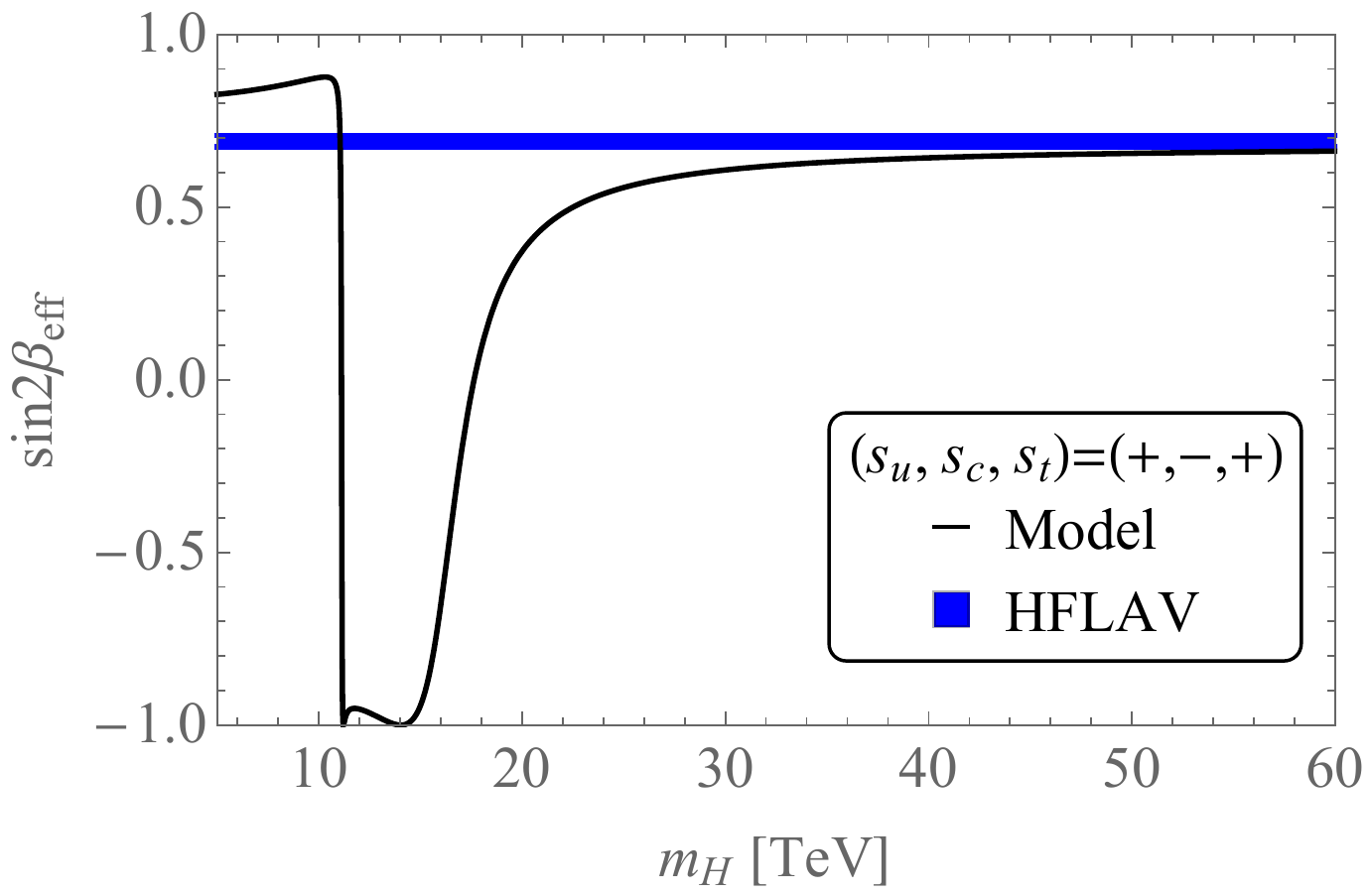}
   \label{Data10}}
   \end{minipage}
\caption{Model prediction for $\sin2\beta_{\mathrm{eff}}$
with (a) $(s_u, s_c, s_t)=(+, +, +)$ and
(b) $(s_u, s_c, s_t)=(+, -, +)$.}
\label{Fig:5}
\end{figure}
\par
In Figure\ \ref{Fig:4},
in order to illustrate the FCNH mass dependence of
$\sin2\beta_{\mathrm{eff}}$, we plot,
\bea
\mathrm{arg}\left(1+
\frac{M_{12}^{d\mathrm{FCNH}}}{M_{12}^{d\mathrm{SM}}}\right),
\label{EQARG}
\eea
where for the CKM matrix components,
we used the values in Eq.\ (\ref{Eq:CKMf}).
The numerical behavior of the argument in Eq.\ (\ref{EQARG})
can be understood in the following way:
If one neglects masses of up and charm quarks,
a phase for $M_{12}^{d\mathrm{FCNH}}$ is determined as
$-(V^\dagger m_u^D V)_{bd}^2\propto -(V_{tb}^* V_{td})^2$.
This factor is the same as the box diagram in the SM up to its sign.
Therefore, $M_{12}^{d \mathrm{FCNH}}/M_{12}^{d \mathrm{SM}}$ is
a negative real number, approximately.
Owing to this fact, we find
\bea
\begin{cases}
\mathrm{arg}(1+M_{12}^{d\mathrm{FCNH}}/M_{12}^{d\mathrm{SM}})=0,\qquad
(-1<M_{12}^{d\mathrm{FCNH}}/M_{12}^{d\mathrm{SM}}<0)\\
\mathrm{arg}(1+M_{12}^{d\mathrm{FCNH}}/M_{12}^{d\mathrm{SM}})=
\pm \pi,
\qquad
(M_{12}^{d\mathrm{FCNH}}/M_{12}^{d\mathrm{SM}}<-1)
\end{cases}\label{Eq:arg}
\eea
where for $M_{12}^{d\mathrm{FCNH}}/M_{12}^{d\mathrm{SM}}<-1$,
sign of the argument depends on sign of tiny imaginary part
in the argument.
The relations in Eq.\ (\ref{Eq:arg})
indicate that 
the argument in Eq.\ (\ref{EQARG}) vanishes
if $m_H$ is sufficiently large.
From Figure\ \ref{Fig:4}, we find that
the argument in Eq.\ (\ref{Eq:arg}) does not affect
$\sin2\beta_{\mathrm{eff}}$ for $m_H\geq 12\ \mathrm{TeV}$.
Note that in fact, charm quark mass slightly contributes to the above quantity so that the exact behavior in Figure\ \ref{Fig:4}
is not identical to Eq.\ (\ref{Eq:arg}).
\par
We should also note that quark mass signs
in Eq.\ (\ref{Eq:mass}) give rise to
numerical difference in
the argument in Eq.\ (\ref{EQARG}).
However, choice of $s_u=\pm 1$ yields
a minor difference since up quark mass is
negligible.
Thus, the two representative cases,
$(s_u, s_c, s_t)=(+, +, +)$ and $(s_u, s_c, s_t)=(+, -, +)$
are sufficient
since the relative sign for $s_c$ and $s_t$
almost determines the coupling for $\Delta F=2$ processes
in Eq.\ (\ref{Cs}).
\par
In Figure\ \ref{Fig:5}, we show the prediction
for $\sin2\beta_{\mathrm{eff}}$
with $(s_u, s_c, s_t)=(+, +, +)$ and
$(s_u, s_c, s_t)=(+, -, +)$.
In the plot, one can find that $\sin2\beta_{\mathrm{eff}}$
shows non-trivial dependence on
Higgs mass since both $K$ and $B_d^0$ systems affect
the observable.
For $m_H\gg 50\ \mathrm{TeV}$,
the neutral Higgs decouples from
$\sin2\beta_{\mathrm{eff}}$
so that it becomes identical
to the SM prediction asymptotically.
\subsection{Bound on $m_H$ from CKM fitting}
Now that we have obtained the formulae for flavor-violating observables,
 we derive bounds on FCNH mass.
\par
It should be noted that in the presence of FCNH, the values of CKM matrix components are altered from those found in the literature,
 since the FCNH exchange process modifies the theoretical formulae for $\Delta M_{B_d}, \Delta M_{B_s}, \mathrm{Re}\ \epsilon$ and $\sin 2\beta_{\mathrm{eff}}$.
In order to derive the bounds,
we utilize p-values in the CKM fitting for flavor-violating observables.
The analysis is performed in the following way:
We carry out $\chi^2$ fittings with fixed $m_H$.
For each value of $m_H$, p-values are given
to specify a disfavored range of Higgs mass.
Given the stringent experimental constraint on $\mathrm{Re}\ \epsilon$,
 the FCNH must be sufficiently heavy.
Thus, $m_H\geq 20\ \mathrm{TeV}$ is considered
in the fittings.
\par
For completeness, we show the Wolfenstein
parametrization of the CKM matrix \cite{Wolfenstein:1983yz},
\bea
V=\begin{pmatrix}
1-\lambda^2/2 & \lambda & A\lambda^3(\rho-
i \eta)\\
-\lambda & 1-\lambda^2/2 & A\lambda^2\\
A\lambda^3(1-\rho-i \eta) & -A\lambda^2 & 1
\end{pmatrix},\label{CKMpara}
\eea
where terms of $\mathcal{O}(\lambda^4)$
are ignored.
In Eq.\ (\ref{CKMpara}), $(\rho, \eta)$
are redefined in terms of
phase convention independent parameters,
$(\bar{\rho}, \bar{\eta})$ \cite{Buras:1994ec},
\bea
\rho+i\eta =\frac{(\bar{\rho}+i\bar{\eta})
\sqrt{1-A^2\lambda^4}}{\sqrt{1-\lambda^2}
[1-A^2\lambda^4(\bar{\rho}+i\bar{\eta})]}.
\eea
In addition, angles which constitute the unitarity triangle
are defined as,
\bea
\alpha&=&\arg\left(-\frac{V_{td}V_{tb}^*}{V_{ud}V_{ub}^*}\right),\\
\gamma&=&\arg\left(-\frac{V_{ud}V_{ub}^*}{V_{cd}V_{cb}^*}\right),\qquad
\eea
where the other angle, $\beta$, is given in Eq.\ (\ref{Eq:beta}).
We note that $(\alpha, \beta, \gamma)$ can be written in terms
of the Wolfenstein parameters, $(\lambda, A, \bar{\rho}, \bar{\eta})$. 
\par
In carrying out the analysis,
the following facts are considered:
\begin{itemize}
\item Measurements of $|V_{ud}|, |V_{us}|, |V_{ub}|,
|V_{cd}|, |V_{cs}|$ and $|V_{cb}|$ are unchanged
in the presence of FCNH.
This is because these absolute values are determined through
semi-leptonic decays, while the model
predicts minor correction for semi-leptonic processes.
\item Angle $\alpha$, which is measured in
time-dependent processes for
$B\to\pi\pi, B\to\pi\rho$ and $B\to\rho\rho$ decays,
is altered by FCNH, since these observables
are sensitive to $B_d^0-\bar{B_d^0}$ mixing.
However, as stated in the previous subsection,
the FCNH contribution to $B_d^0-\bar{B_d^0}$ mixing does not affect $q/p$
for $m_H\geq 12\ \mathrm{TeV}$
so that we decouple this effect.
\item Measurement of angle $\gamma$ is carried out in
$B^{\pm}\to DK^{\pm}$ and $B^{0}\to D^{(*)\pm}\pi^{\mp}$ 
decays.
These tree-level processes are not significantly modified by
the FCNH exchange.
\end{itemize}
\par
In the fitting, we include uncertainties of 
the Wolfenstein parameters \cite{Charles:2015gya},
the bag parameters
\cite{Carrasco:2015pra, Carrasco:2013zta}
and NLO QCD correction factors
for the $\Delta S=2$ process denoted by $\eta_1$,
$\eta_2$ and $\eta_3$ \cite{Herrlich:1993yv}.
As for the $\Delta B=2$ processes,
uncertainty in short-distance QCD correction factors
is not considered because they are more precisely determined
than those for the $\Delta S=2$ process.
We use the central value of this QCD correction calculated through the method of Ref.\ \cite{Buras:2001ra}.
\par
In the CKM fitting, a statistic,
\bea
(\chi^2)_{\mathrm{fixed}\ m_H}&=&\displaystyle\sum_{i, j}
\frac{(|V_{ij}|_{\mathrm{Th}}-|V_{ij}|_{\mathrm{Exp}})^2}{(\sigma_{|V_{ij}|})^2_{\mathrm{Exp}}}+
\displaystyle\sum_{k={\alpha, \gamma, \sin2\beta_{\mathrm{eff}}}}
\frac{(k_{\mathrm{Th}}-k_{\mathrm{Exp}})^2}{(\sigma_{k})_{\mathrm{Exp}}^2}
\nn\\
&&
+\displaystyle\sum_{l=d, s}
\frac{(\Delta M_{B_l \mathrm{Th}}-
\Delta M_{B_l\mathrm{Exp}})^2}{
(\sigma_{\Delta M_{B_l}})_{\mathrm{Th}}^2+
(\sigma_{\Delta M_{B_l}})_{\mathrm{Exp}}^2}
+\frac{(\mathrm{Re} \epsilon_{\mathrm{Th}}-\mathrm{Re} \epsilon_{\mathrm{Exp}})^2}
{(\sigma_{\mathrm{Re} \epsilon})^2_{\mathrm{Th}}+
(\sigma_{\mathrm{Re} \epsilon})^2_{\mathrm{Exp}}},
\label{Eq:chisq}
\eea
is minimized with
$(i, j)=(u, d), (u, s), (c, d), (c, s), (c, b), (u, b)$.
Experimental data on the absolute values of the CKM matrix elements 
and $\alpha, \gamma$ are provided by PDG as,
\bea
|V_{ud}|_{\mathrm{Exp}}&=&0.97417\pm 0.00021,\nn\\
|V_{us}|_{\mathrm{Exp}}&=&0.2248\pm 0.0006,\nn\\
|V_{cd}|_{\mathrm{Exp}}&=&0.220\pm0.005,\nn\\
|V_{cs}|_{\mathrm{Exp}}&=&0.995\pm 0.016,\nn\\
|V_{cb}|_{\mathrm{Exp}}&=&(40.5 \pm 1.5) \times 10^{-3},\nn\\
|V_{ub}|_{\mathrm{Exp}}&=&(4.09 \pm 0.39) \times 10^{-3},\nn\\
\alpha_{\mathrm{Exp}}&=&(87.6\pm3.4)^\circ,\nn\\
\gamma_{\mathrm{Exp}}&=&(73.2\pm6.7)^\circ,\nn
\eea
where we have taken averages of errors for $\alpha_{\mathrm{Exp}}$
and $\gamma_{\mathrm{Exp}}$,
which are originally given as asymmetric forms.
The data of $(\sin2\beta_{\mathrm{eff}})_{\mathrm{Exp}},
(\mathrm{Re}\ \epsilon)_{\mathrm{Exp}}, (\Delta M_{B_d})_{\mathrm{Exp}},
(\Delta M_{B_s})_{\mathrm{Exp}}$
are extracted from Table\ \ref{Tab:1}.
In the r.h.s of Eq.\ (\ref{Eq:chisq}),
note that Higgs mass is fixed and the Wolfenstein parameters are adjustable to minimize $(\chi^2)_{\mathrm{fixed}\ m_H}$.
Furthermore, in Eq.\ (\ref{Eq:chisq}), the theoretical errors 
for the bag parameters
in $\Delta M_{B_d}, \Delta M_{B_s}$
are added to experimental ones in quadrature
while for $\mathrm{Re}\ \epsilon$, the errors of
perturbative QCD factors $(\eta_1, \eta_2, \eta_3)$
are also accounted.
Note that our fitting analysis is different from the
one performed by the CKMfitter group
\cite{Charles:2015gya},
which is based on {\it R}fit \cite{Hocker:2001xe},
another frequentist approach to include
theoretical uncertainty.\par
In Figure\ \ref{Fig:6}, p-values obtained for $20\ \mathrm{TeV}\leq m_H\leq 350\ \mathrm{TeV}$
are presented.
One observes a difference between $s_u s_c=+1$
and $s_u s_c=-1$.
From Figure\ \ref{Fig:6}, we derive lower bounds on the FCNH mass
in which p-values are disfavored by $3\sigma$ and $5\sigma$.
For the two cases of sign choice,
the bounds are
\bea
&&\mathrm{For}\ (s_u, s_c, s_t)=(+, +, +),\qquad
m_H > 84\ \mathrm{TeV}\ (3\sigma),\qquad 
m_H > 64\ \mathrm{TeV}\ (5\sigma),\\
&&\mathrm{For}\ (s_u, s_c, s_t)=(+, -, +),\qquad
m_H > 75\ \mathrm{TeV}\ (3\sigma),\qquad
m_H > 56\ \mathrm{TeV}\ (5\sigma).
\eea
\begin{figure}[h!]
\includegraphics[width=12.2cm]{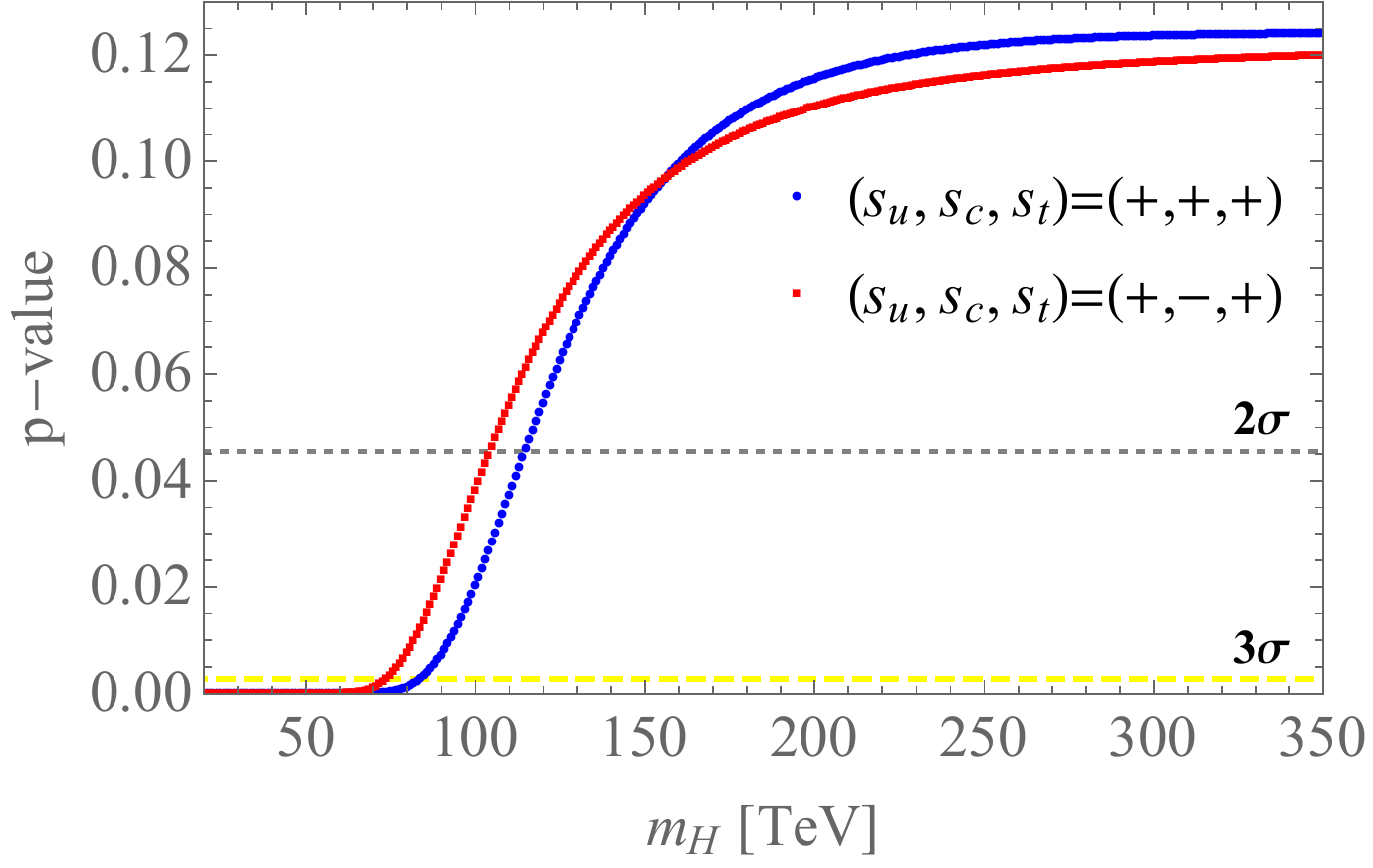}
   \label{Data13}
\caption{P-value obtained in CKM fitting.
Blue dots and red rectangles represent
the cases of $(s_u, s_c, s_t)=(+, \pm, +)$.
A gray dotted line and a yellow dashed line
stand for $(100-95.4)\%$ and $(100-99.7)\%$, which correspond to
p-values of $2\sigma$ and $3\sigma$ confidence levels
(CLs), respectively.}
\label{Fig:6}
\end{figure}
\section{Prediction for
$\mathrm{Re}\ \epsilon, \Delta M_{B_d}$ and $\Delta M_{B_s}$}
In this section, we present the prediction for observables
to illustrate the pattern of deviation in the model.
For this purpose, the Wolfenstein parameters are estimated
from observables which are not affected by FCNH.
Hence, we consider the following statistic,
\bea
\chi^2&=&\displaystyle\sum_{i, j}
\frac{(|V_{ij}|_{\mathrm{Th}}-|V_{ij}|_{\mathrm{Exp}})^2}{(\sigma_{|V_{ij}|})^2_{\mathrm{Exp}}}+
\frac{(\gamma_{\mathrm{Th}}-\gamma_{\mathrm{Exp}})^2}{(\sigma_{\gamma})_{\mathrm{Exp}}^2}.
\label{Eq:chisq2}
\eea
Through the minimization of Eq.\ (\ref{Eq:chisq2}),
one can extract $(\lambda, A, \bar{\rho}, \bar{\eta})$.
\par
The Belle II experiment announces that
the expected integrated luminosity
is $50\ \mathrm{ab}^{-1}$ in five years of running.
Motivated by this, we consider Case I and Case II described below.

\begin{figure}[h!]
  \setlength{\subfigwidth}{.5\linewidth}
  \addtolength{\subfigwidth}{-.5\subfigcolsep}
  \begin{minipage}[b]{\subfigwidth}
    \subfigure[]{\includegraphics[width=8.2cm]{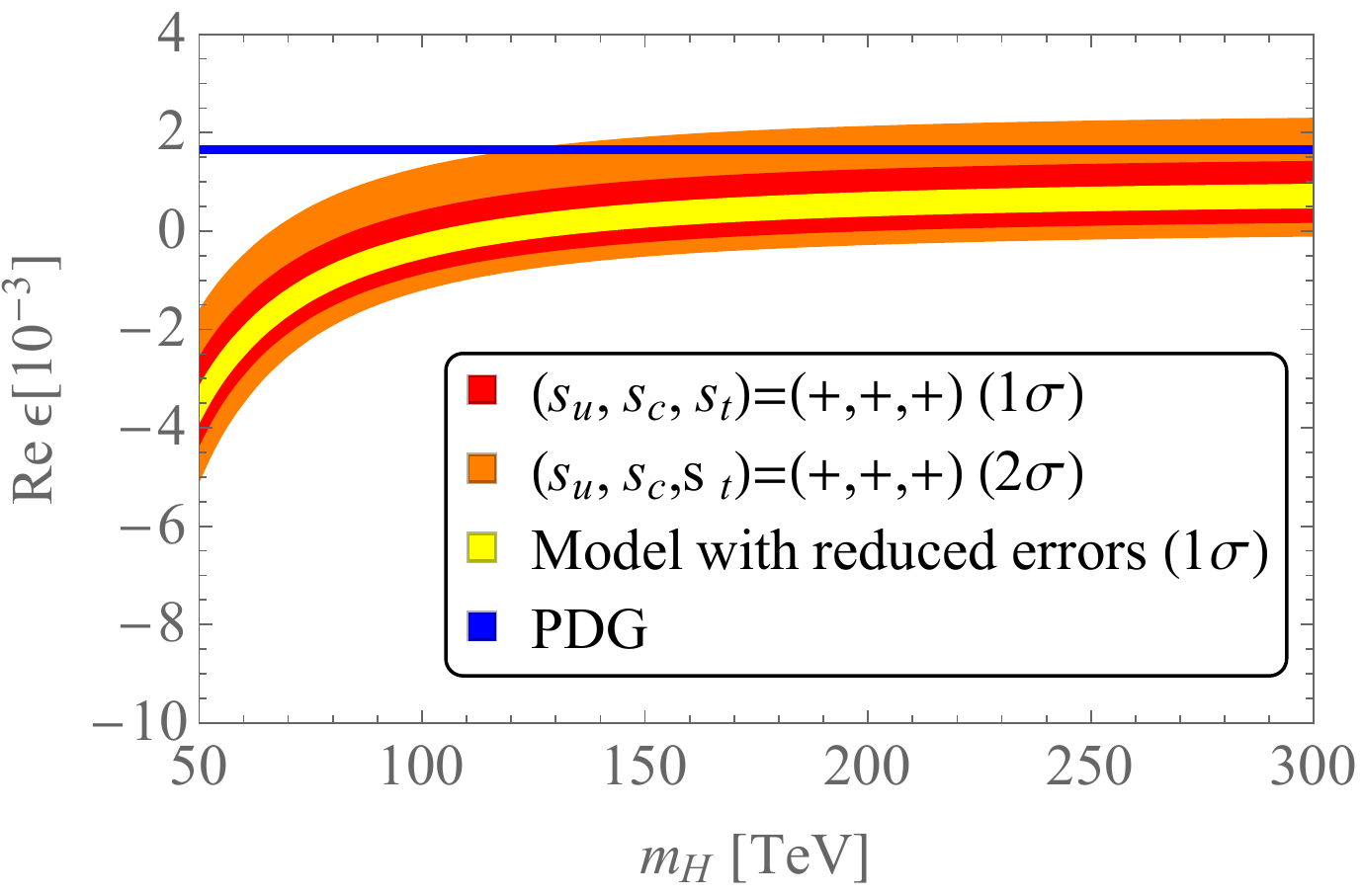}
   \label{Data1}}
  \end{minipage}
  \begin{minipage}[b]{\subfigwidth}
    \subfigure[]{\includegraphics[width=8.2cm]{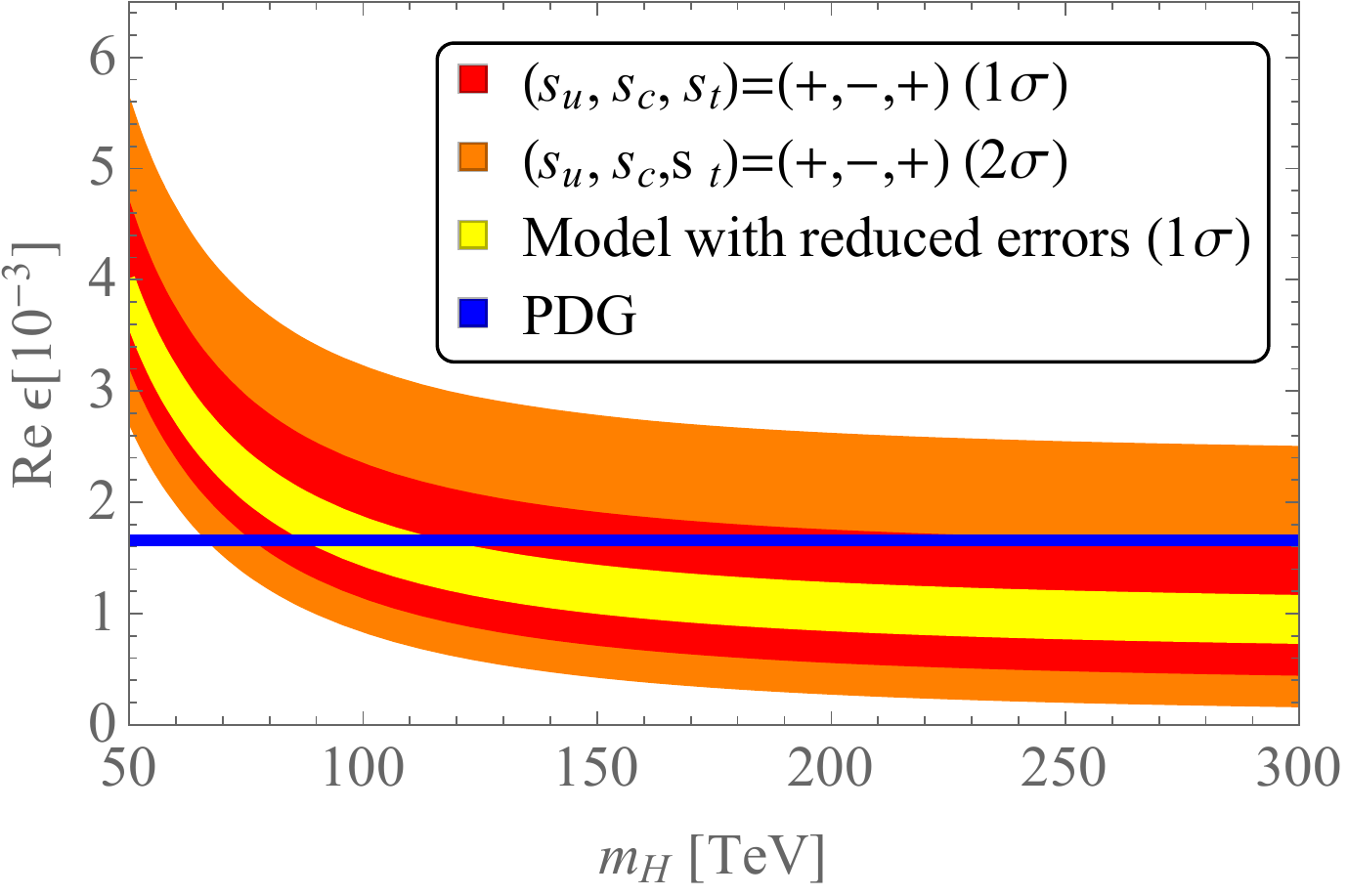}
   \label{Data2}}
  \end{minipage}\\
    \begin{minipage}[b]{\subfigwidth}
    \subfigure[]{\includegraphics[width=8.2cm]{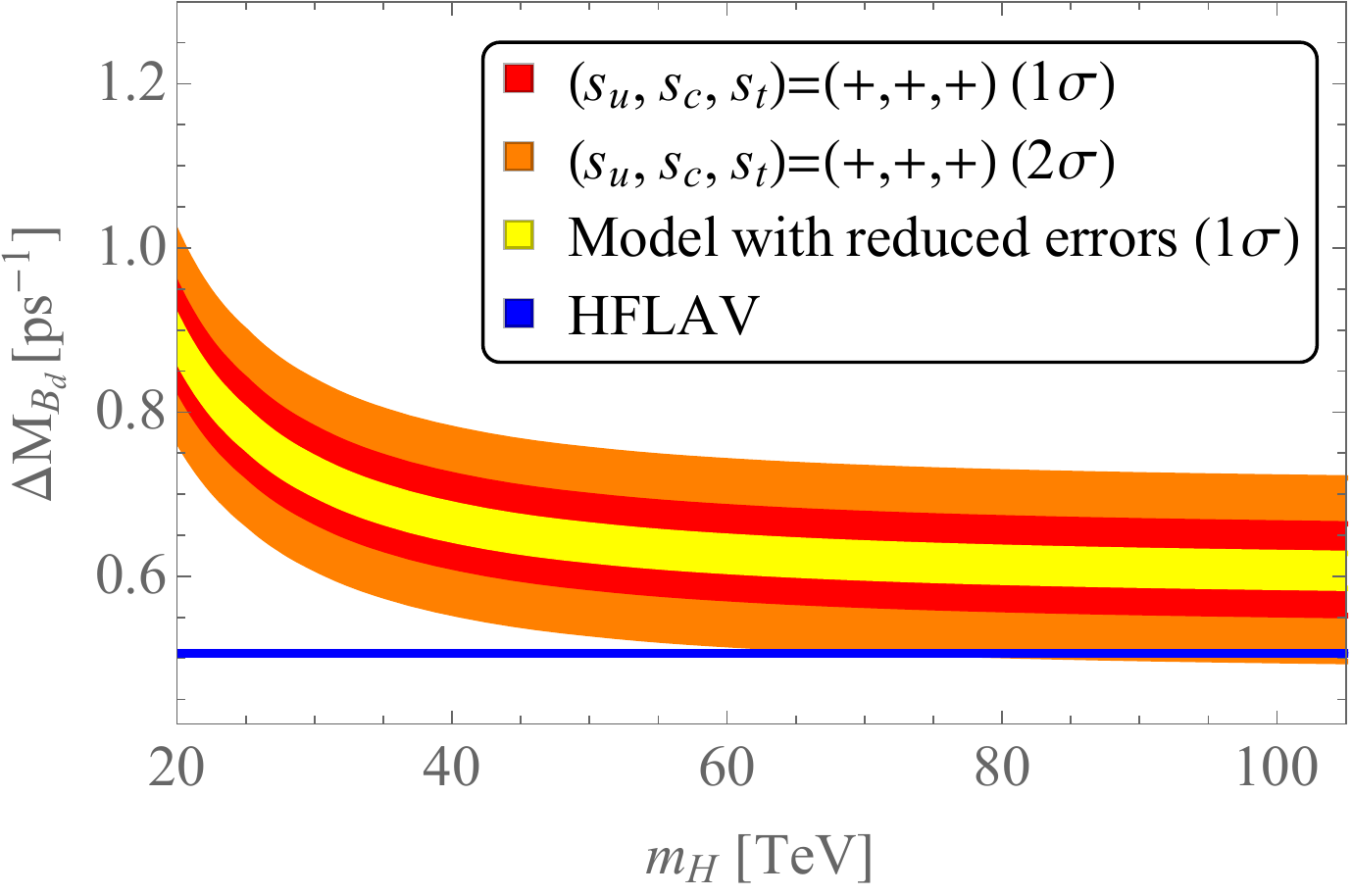}
   \label{Data3}}
  \end{minipage}
      \begin{minipage}[b]{\subfigwidth}
    \subfigure[]{\includegraphics[width=8.2cm]{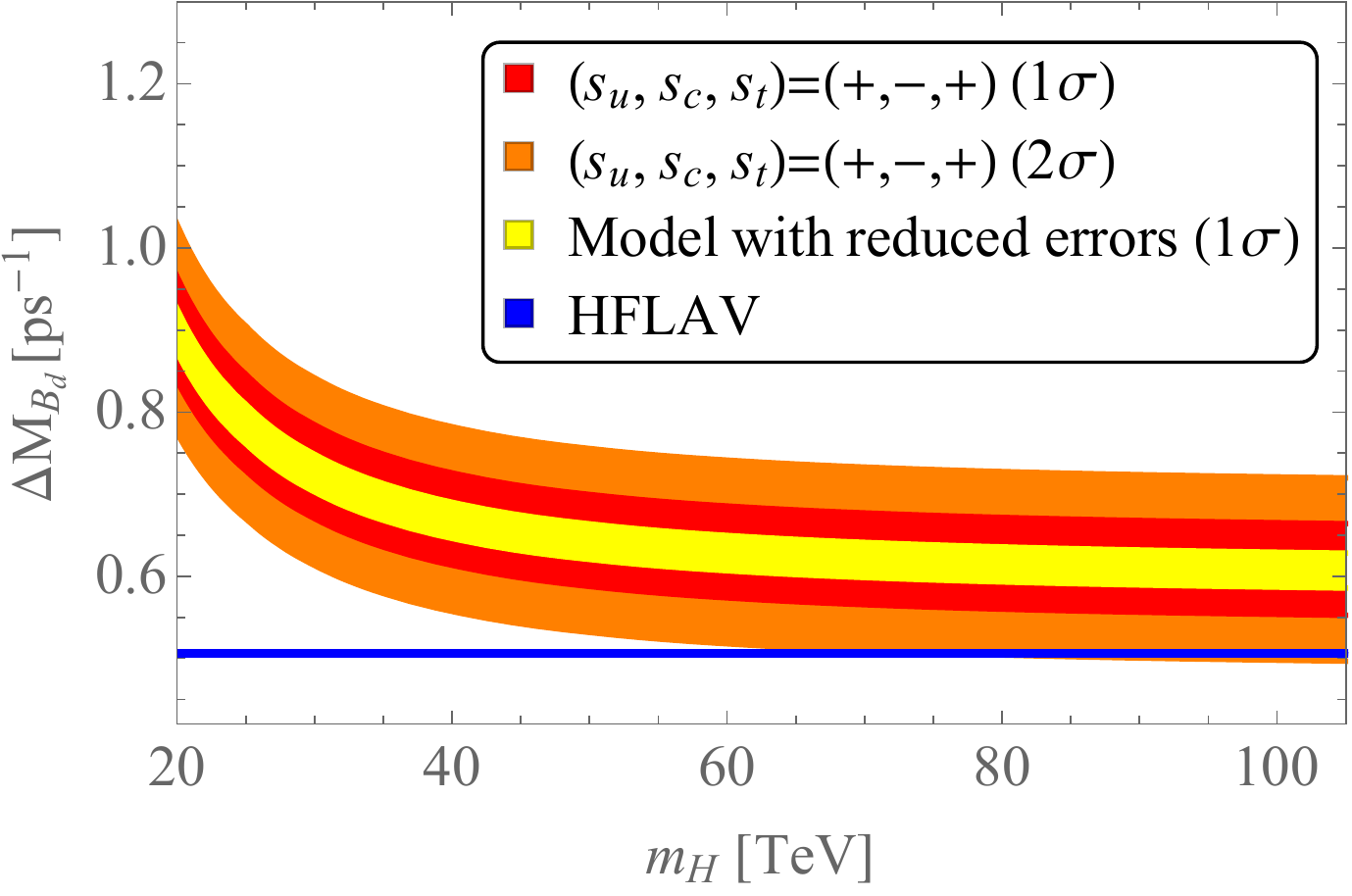}
   \label{Data3}}
   \end{minipage}
       \begin{minipage}[b]{\subfigwidth}
    \subfigure[]{\includegraphics[width=8.2cm]{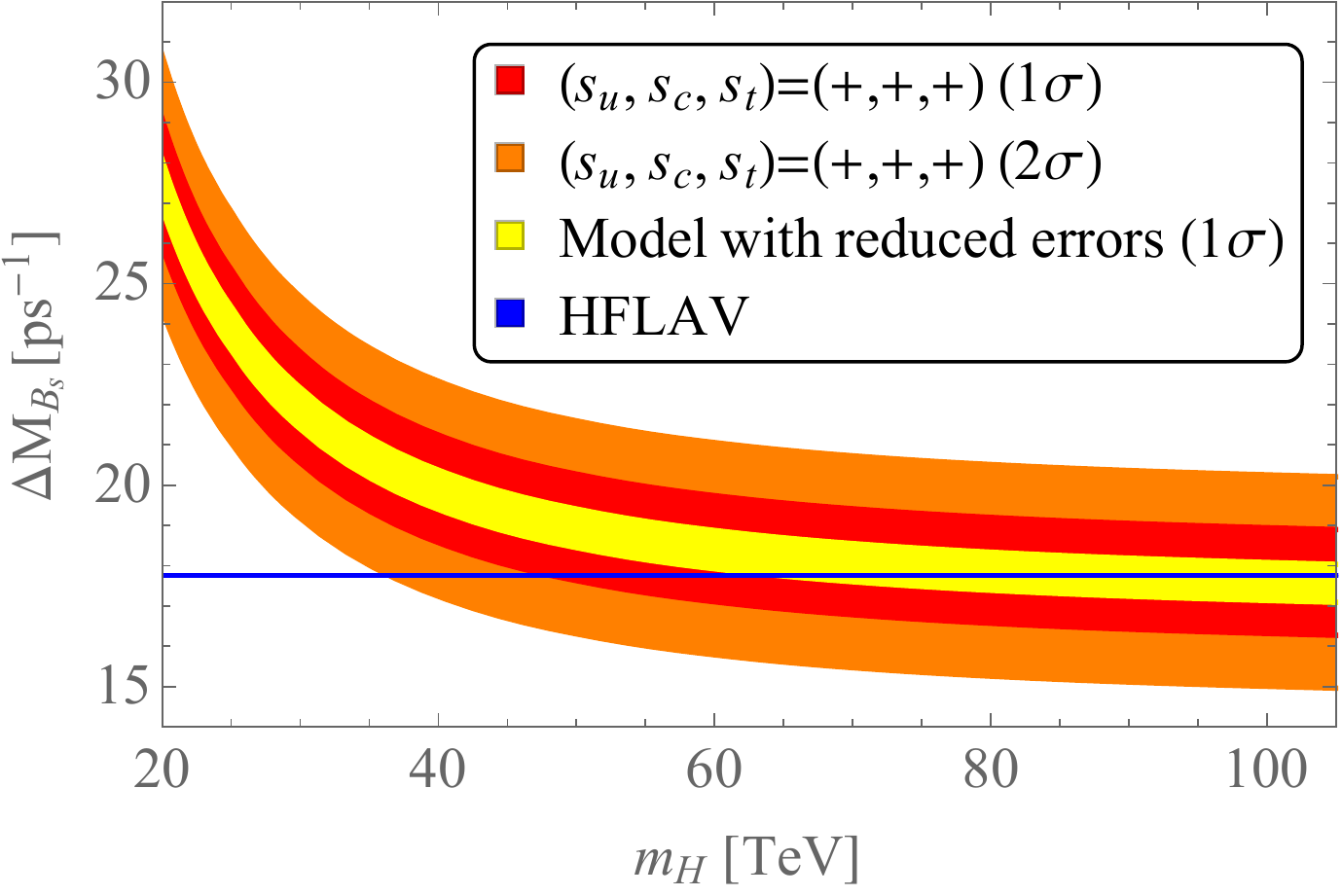}
   \label{Data3}}
  \end{minipage}
      \begin{minipage}[b]{\subfigwidth}
    \subfigure[]{\includegraphics[width=8.2cm]{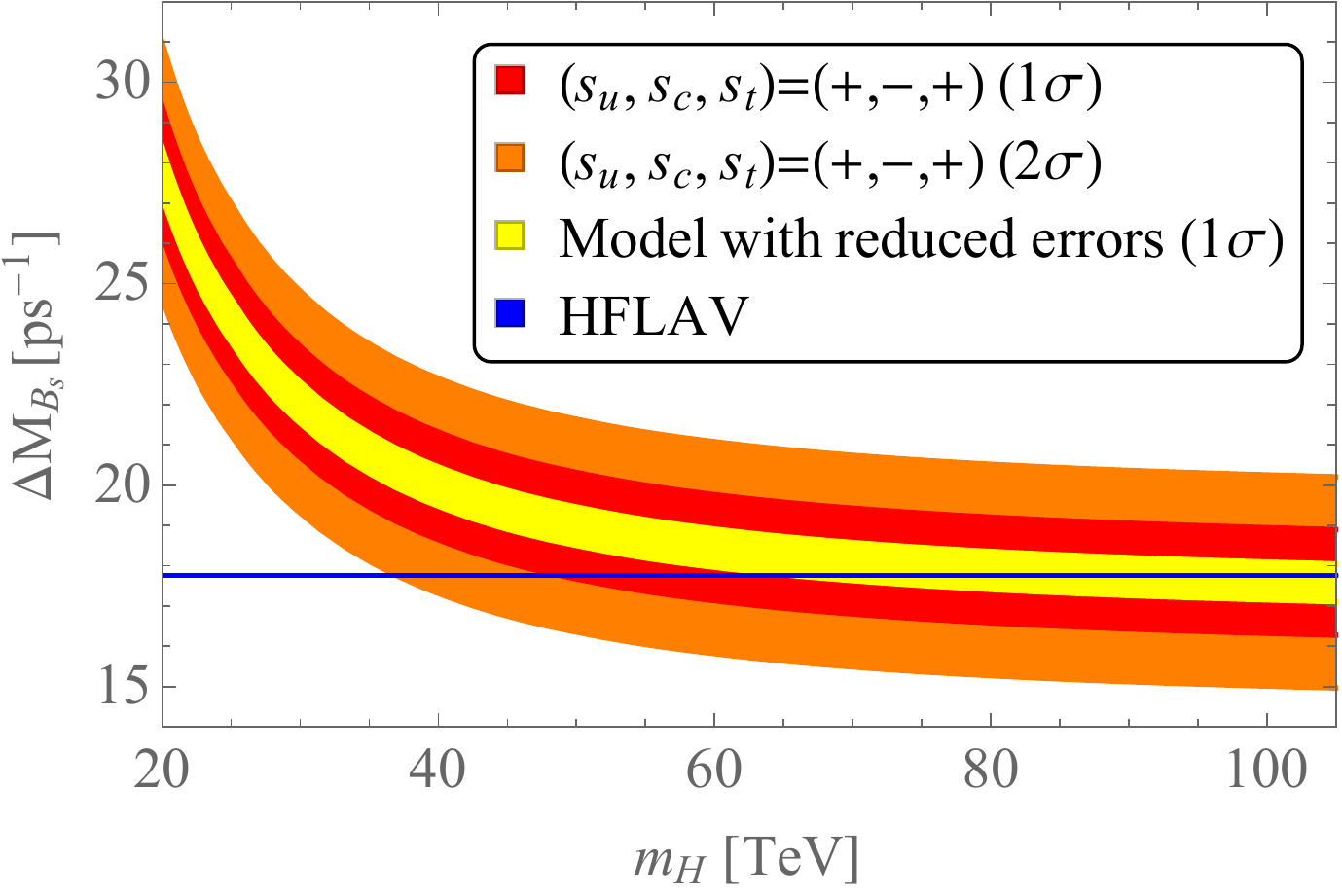}
   \label{Data3}}
  \end{minipage}
\caption{
Same figure as Fig.\ \ref{Fig:6} in Case II.
}
\label{Fig:7}
\end{figure}

\begin{itemize}

\item Case I:

The errors of $|V_{ub}|$ and $\gamma$ are reduced by
$1/3$ and $1/7$, respectively, without changing their central values.
The errors and central values of the other quantities, including $|V_{cb}|$, remain the same.
Under this circumstance,
the parameters are estimated as
\bea
&\lambda=0.22547\pm0.00050,\quad
A=0.797\pm0.030,&\nn\\
&\bar{\rho}=0.1262\pm0.093,\quad
\bar{\eta}=0.418\pm0.021,&\label{Eq:estimated}
\eea
where a correlation matrix for $(\lambda, A, \bar{\rho},
\bar{\eta})$ is
\bea
\begin{pmatrix}
1 &\hspace{4.5mm} -0.12 &\hspace{2mm} -0.032 &\hspace{1.5mm} -0.048\\
* & \hspace{4.5mm}1 & \hspace{2mm}-0.50 &\hspace{1.5mm} -0.74\\
* & \hspace{4.5mm}* & \hspace{2mm}1 &\hspace{1.5mm} 0.59\\
* & \hspace{4.5mm}* & \hspace{2mm}* &\hspace{1.5mm} 1\\
\end{pmatrix}.
\eea
On the basis of the Wolfenstein parameters in Eq.\ (\ref{Eq:estimated}),
the prediction for $(\mathrm{Re}\ \epsilon, \Delta M_{B_d}, \Delta M_{B_s})$
is given in Figure\ \ref{Fig:6}.
We also present the result
for the case in which errors of bag parameters
and decay constants are three times as small as
the ones calculated by the ETM collaboration.
\par

\item Case II:

Case II is an ideal situation in which the errors of $|V_{cb}|, |V_{ub}|$ and $\gamma$
are reduced by $1/7$ (the other quantities remain the same).
In this case, the Wolfenstein parameters
and their correlation matrix are,
\bea
&\lambda=0.22547\pm0.00050,\quad
A=0.7967\pm0.0055,&\nn\\
&\bar{\rho}=0.1262\pm0.071,\quad
\bar{\eta}=0.4180\pm0.0065,&\label{Eq:estimated}\\
&\begin{pmatrix}
1 &\hspace{4.5mm} -0.65 &\hspace{2mm} -0.042 &\hspace{1.5mm} -0.15\\
* & \hspace{4.5mm}1 & \hspace{2mm}-0.044 &\hspace{1.5mm} -0.16\\
* & \hspace{4.5mm}* & \hspace{2mm}1 &\hspace{1.5mm} -0.063\\
* & \hspace{4.5mm}* & \hspace{2mm}* &\hspace{1.5mm} 1\\
\end{pmatrix}.&
\eea
For Case II, the prediction for $(\mathrm{Re}\ \epsilon, \Delta M_{B_d}, \Delta M_{B_s})$
is presented in Figure\ \ref{Fig:7}.
\end{itemize}
\begin{figure}[h!]
  \setlength{\subfigwidth}{.5\linewidth}
  \addtolength{\subfigwidth}{-.5\subfigcolsep}
  \begin{minipage}[b]{\subfigwidth}
    \subfigure[]{\includegraphics[width=8.2cm]{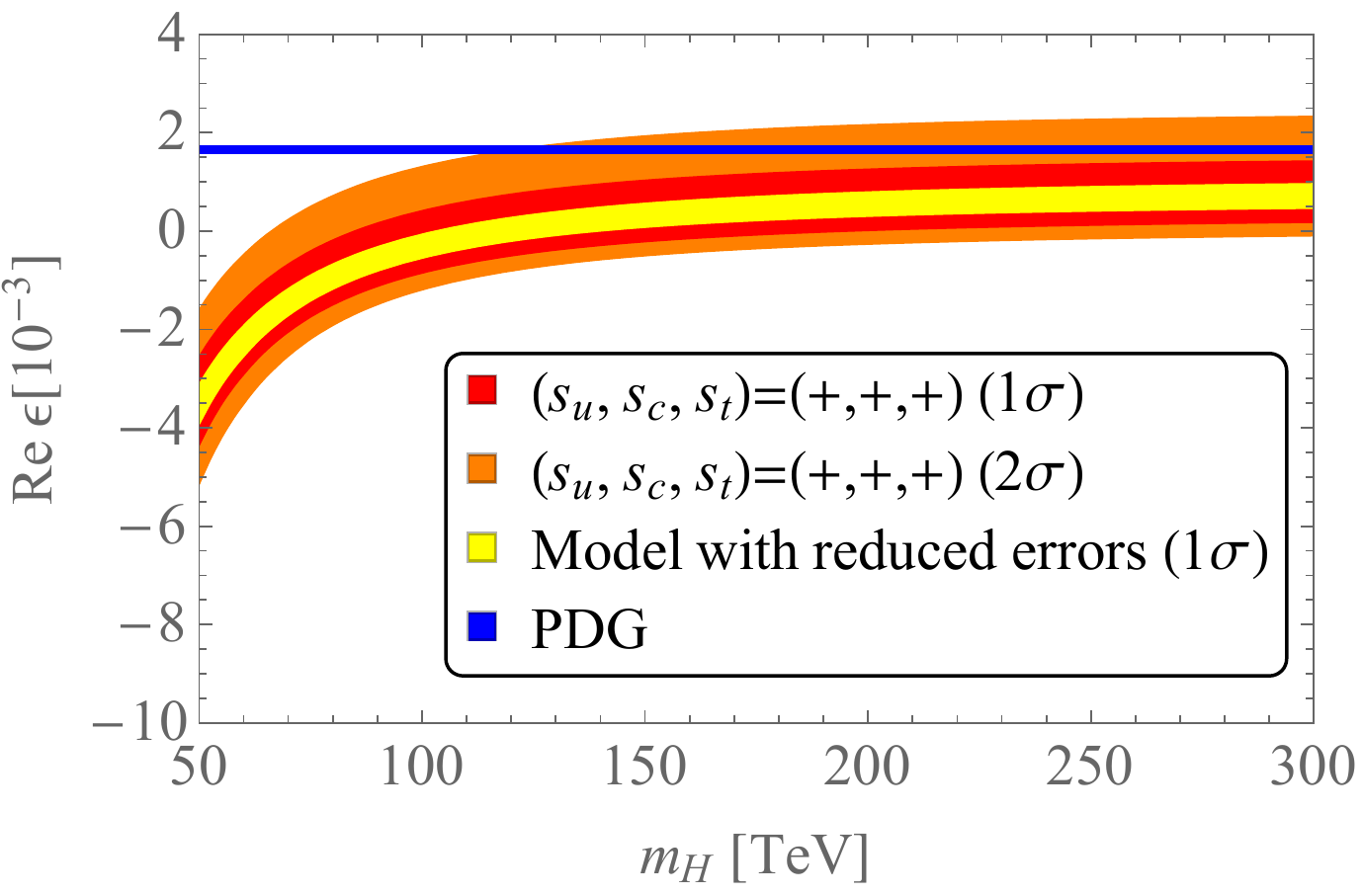}
   \label{Data1}}
  \end{minipage}
  \begin{minipage}[b]{\subfigwidth}
    \subfigure[]{\includegraphics[width=8.2cm]{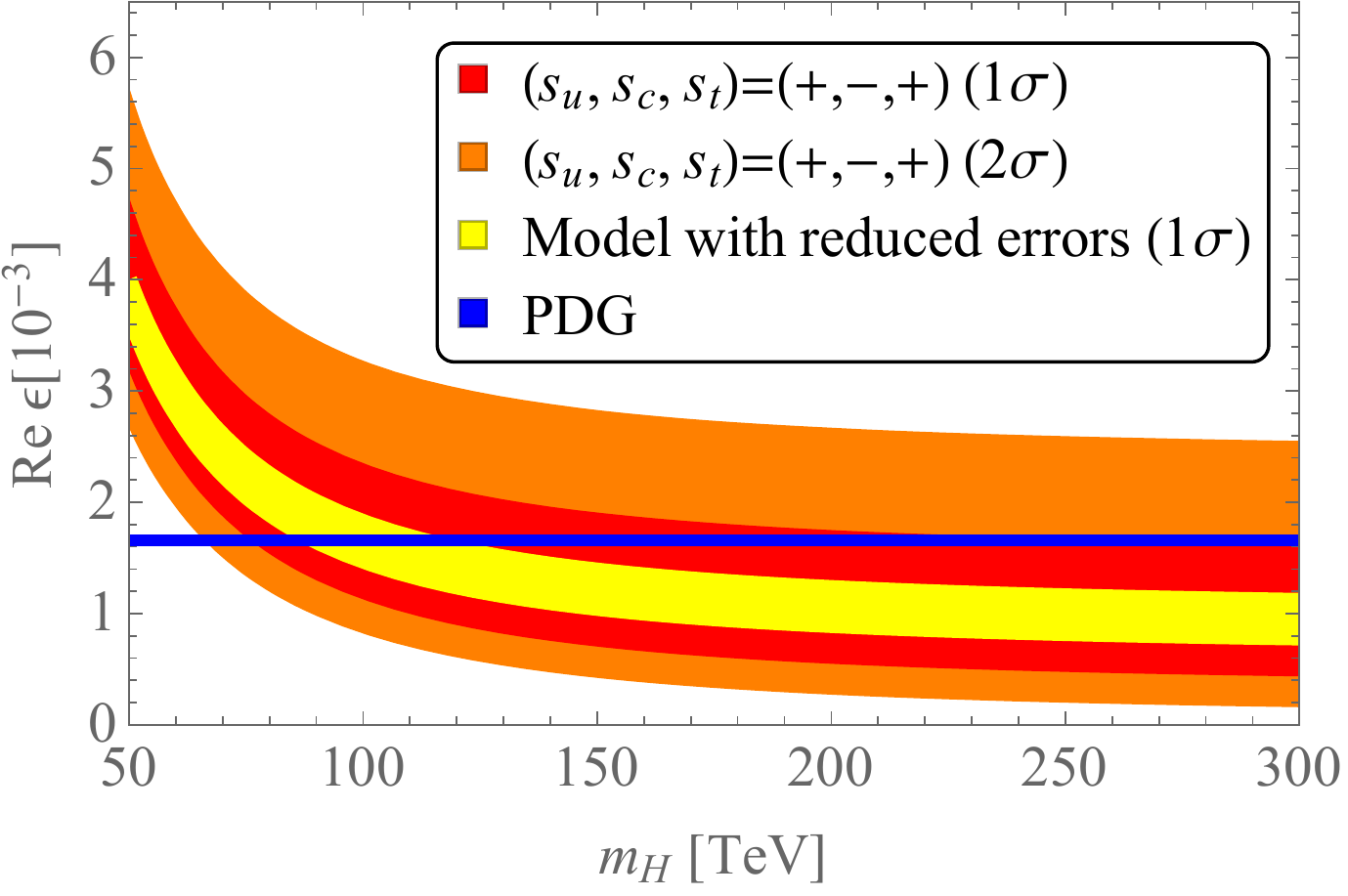}
   \label{Data2}}
  \end{minipage}\\
    \begin{minipage}[b]{\subfigwidth}
    \subfigure[]{\includegraphics[width=8.2cm]{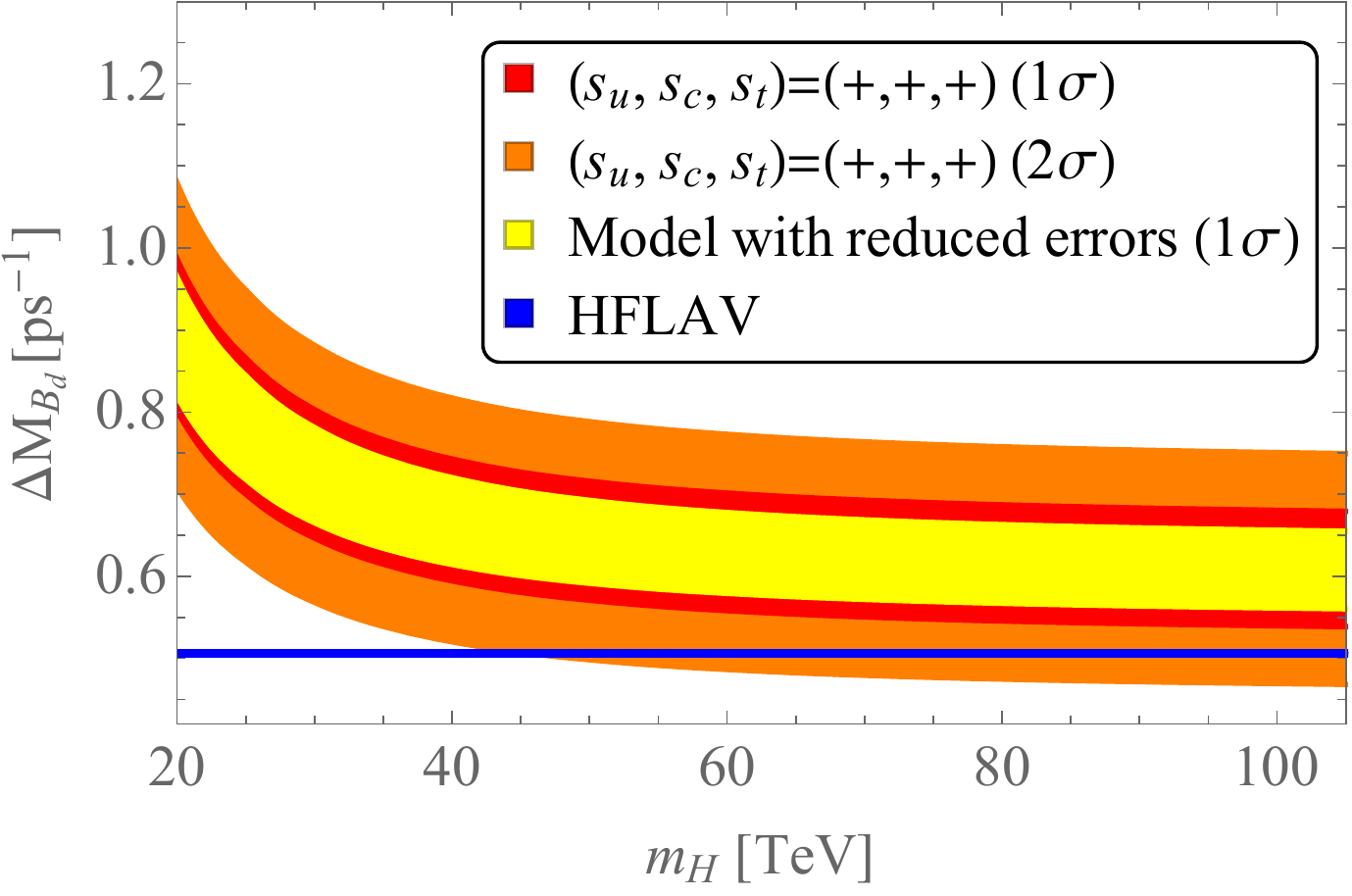}
   \label{Data3}}
  \end{minipage}
      \begin{minipage}[b]{\subfigwidth}
    \subfigure[]{\includegraphics[width=8.2cm]{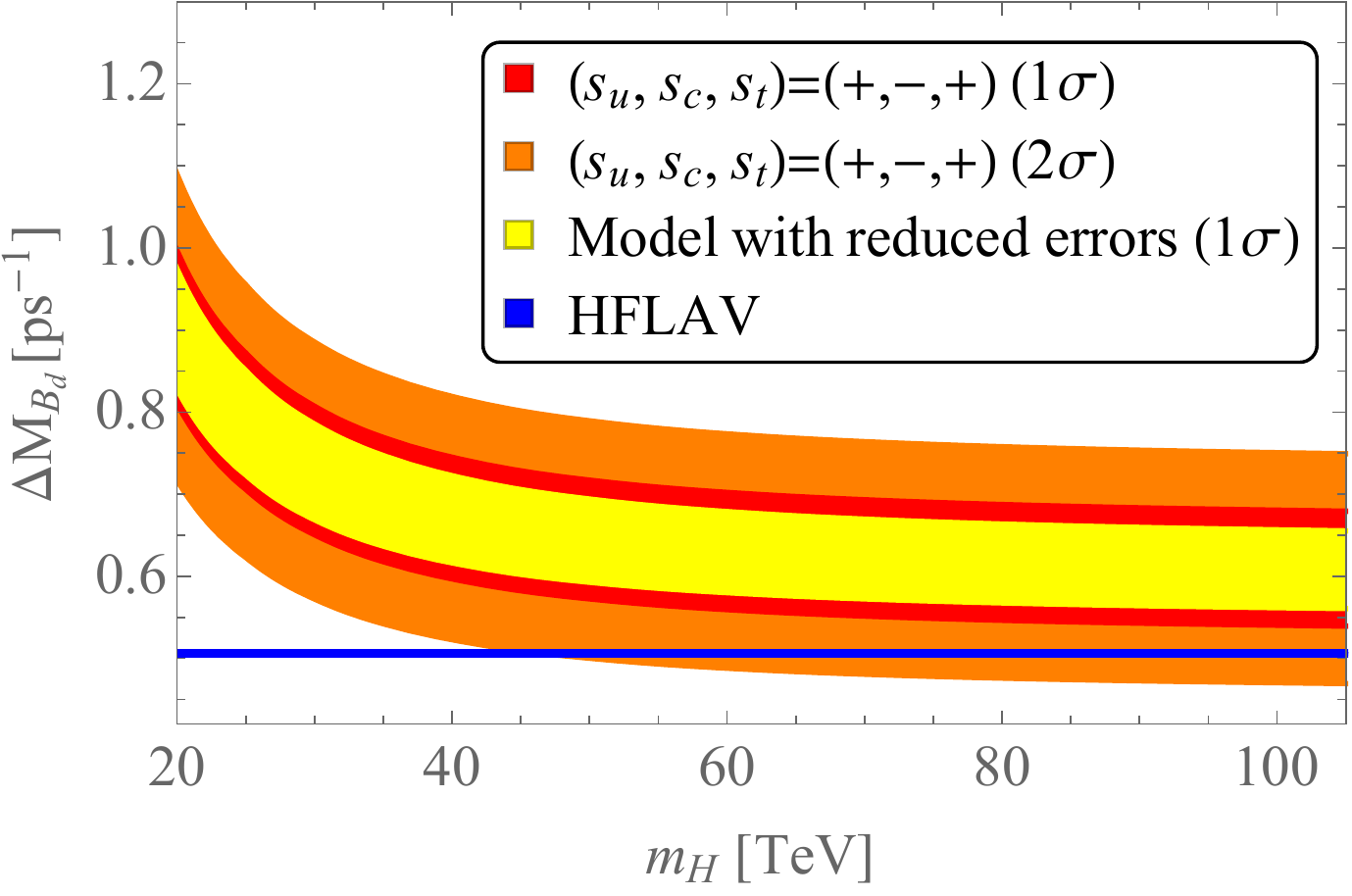}
   \label{Data3}}
   \end{minipage}
       \begin{minipage}[b]{\subfigwidth}
    \subfigure[]{\includegraphics[width=8.2cm]{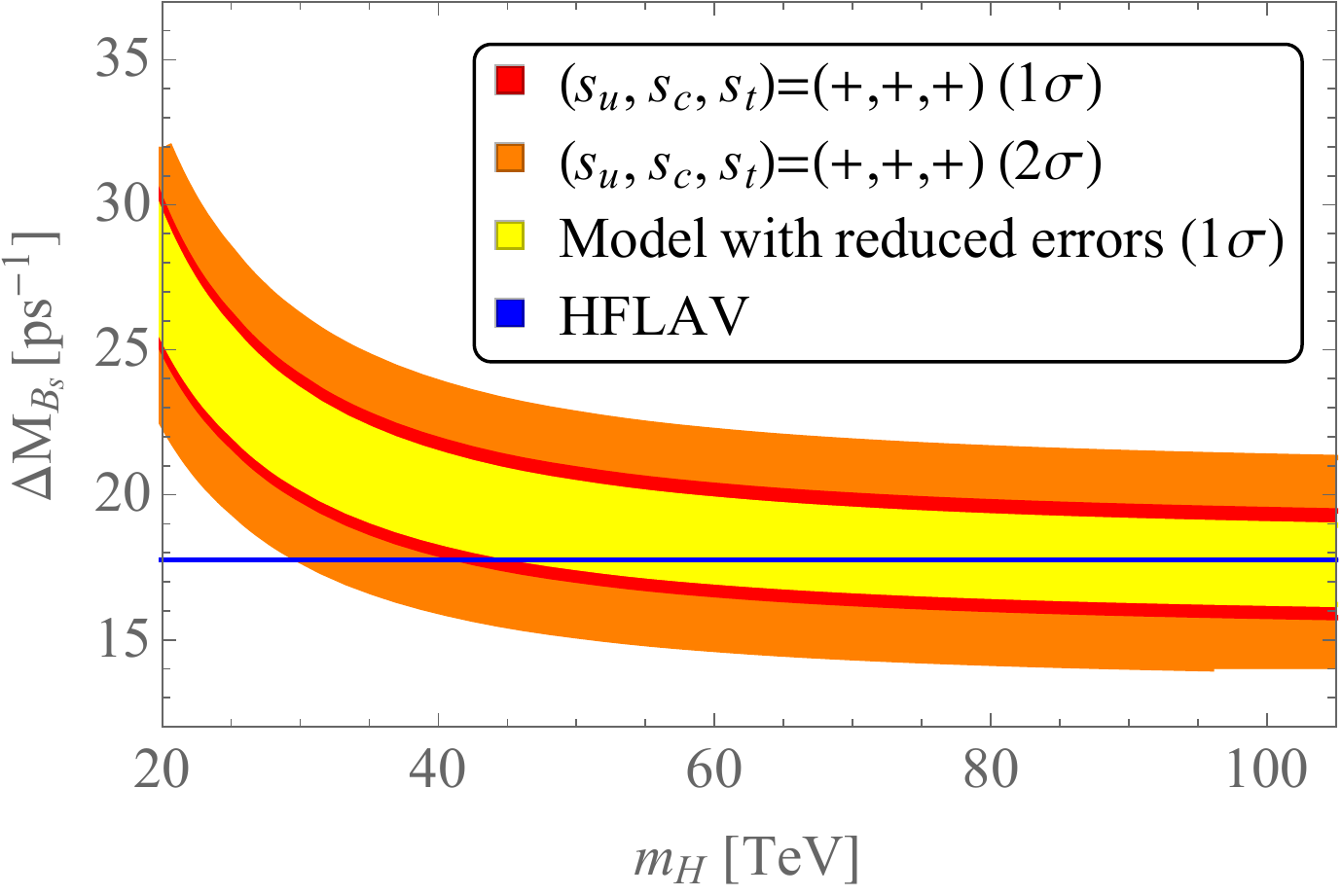}
   \label{Data3}}
  \end{minipage}
      \begin{minipage}[b]{\subfigwidth}
    \subfigure[]{\includegraphics[width=8.2cm]{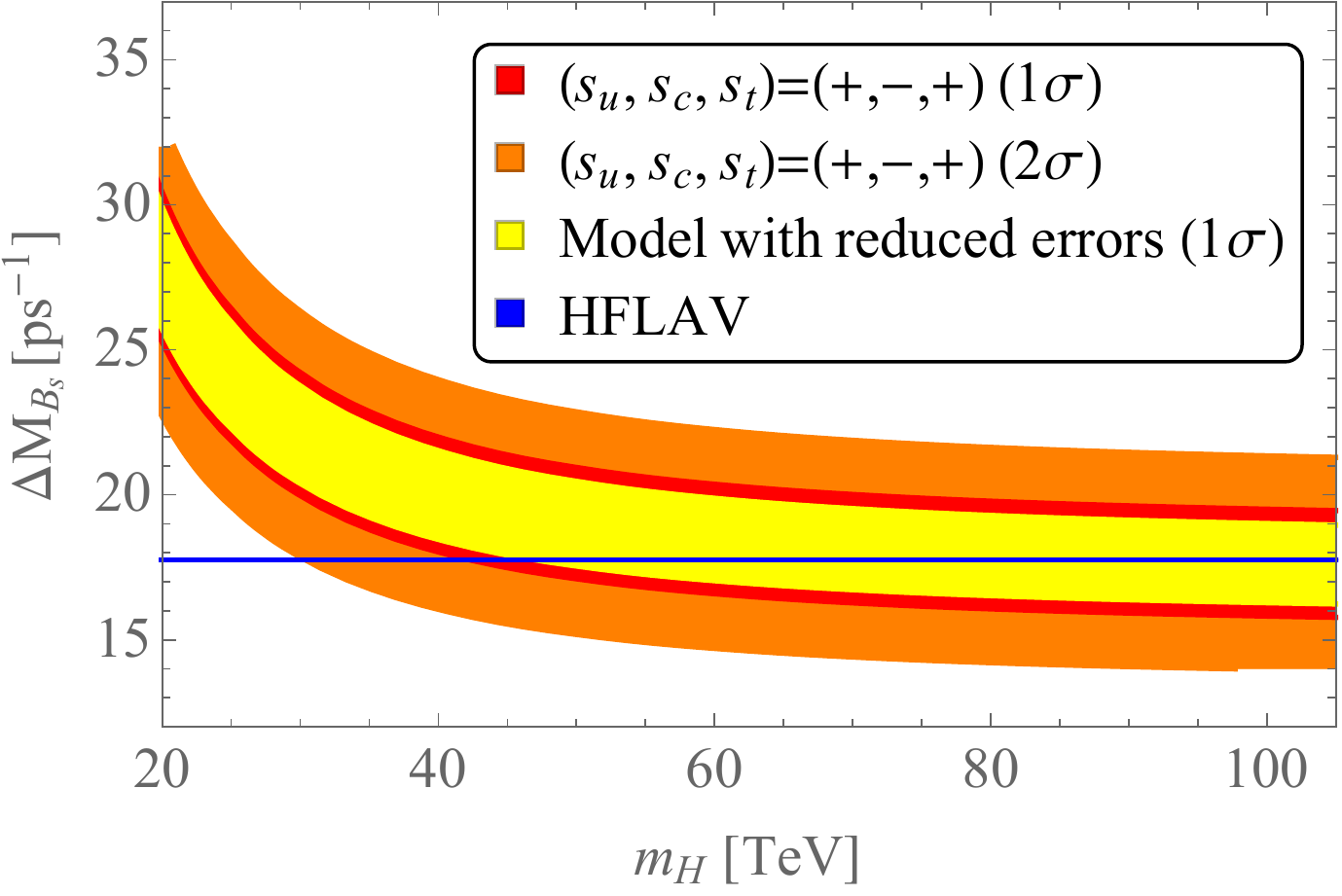}
   \label{Data3}}
  \end{minipage}
\caption{Model predictions for $\mathrm{Re}\ \epsilon$
and $\Delta M_{B_d}, \Delta M_{B_s}$ in Case I.
In these plots, the Wolfenstein parameters used
are estimated through observables
which are not affected by FCNH.
Red (orange) bands represent
the model predictions in $1\sigma$ $(2\sigma)$
CL.
Yellow bands stand for the model predictions
in which the errors of decay constant or bag parameters
are three times as small as the one given by the ETM collaboration.
For comparison, the experimental data of PDG and HFLAV
are shown.}
\label{Fig:6}
\end{figure}
\par
With future precision of
input data and reduction of theoretical uncertainty
as in Case II,
$\mathrm{Re}\ \epsilon, \Delta M_{B_d}$
and $\Delta M_{B_s}$ computed in the SM
 may deviate from the experimental values.
Under this circumstance, it can be the case that theoretical calculations including
the FCNH contributions are consistent with the experimental values.
For example, suppose that the input parameters for $\mathrm{Re}\ \epsilon$
are precisely determined
without changing their central values.
A region of interest is then
$m_H\sim 100\ \mathrm{TeV}$ and $s_cs_t=-1$,
because
the central value of $\mathrm{Re}\ \epsilon$ is consistent with
the experimental value in Figure\ \ref{Fig:6}.
If the model prediction for $\Delta M_{B_d}$ and $\Delta M_{B_s}$ should converge to the current central values,
 then, for $m_H=100\ \mathrm{TeV}$ and $s_cs_t=-1$,
 the prediction and the experimental value of $\Delta M_{B_d}$ would be separated by more than $2\sigma$
 whereas the prediction for $\Delta M_{B_s}$ would be consistent with the experiment.
However, it is possible that theoretical calculations of
$\Delta M_{B_d}$ and $\Delta M_{B_s}$ converge to different values, independently of each other and of
$\mathrm{Re}\ \epsilon$.
If the theoretical inputs for $B_d^0-\bar{B_d^0}$
and $B_s^0-\bar{B_s^0}$
are determined as
\bea
f_{B_d}\sqrt{B_1^{\mathrm{d}VLL}}(m_b)=158.5\ \mathrm{MeV},\qquad
f_{B_s}\sqrt{B_1^{\mathrm{s}VLL}}(m_b)=211.9\ \mathrm{MeV},
\eea
$\Delta M_{B_d}$ and $\Delta M_{B_s}$ will be in agreement with the experiment.
In this case, the deviation between the measurements and calculations of $\mathrm{Re}\ \epsilon$, $\Delta M_{B_d}$ and $\Delta M_{B_s}$ hints at the semi-aligned 2HDM. 

\clearpage
\par
In the following, we comment on other flavor-violating observables.
Since a charged scalar exchange alters
$b\to c l^-\nu$ decay rate at tree level,
this model might be able to address the anomaly in $R_{D^{(*)}}$.
However, this is not the case because
the absolute value of the
$bcH^{-}$ coupling is small up to
$\sqrt{2}|V_{cb}m_b|/v\sim O(10^{-3})$,
which does not exceed the SM $bcW^{-}$ coupling, $|V_{cb}|\sim \mathcal{O}(10^{-2})$.\par
We examine the correction to $b\to ss\bar{d}$ decay.
In the SM, this proceeds via the box diagram and is highly suppressed by
the Glashow-Iliopoulos-Maiani mechanism as
$\mathrm{Br}^{\mathrm{SM}}[b\to ss\bar{d}]=\mathcal{O}(10^{-12})$
\cite{Huitu:1998vn, Lu:2016pfs}.
For the corresponding exclusive mode,
 experimental searches have been performed
\cite{Abbiendi:1999st}
and recently, the LHCb collaboration has reported $\mathrm{Br}[B^-\to K^-K^-\pi^+]<1.1\times10^{-8}\
(90\%\ \mathrm{CL})$
\cite{LHCb:2016rul}.
In Ref.\ \cite{Huitu:1998pa}, 
 the FCNH contribution to the decay width is calculated in the Type-III 2HDM and is found to be
\bea
\Gamma^{\mathrm{FCNH}}[b\to ss\bar{d}]=
\frac{m_b^5}{3072(2\pi)^3}
\left|\displaystyle\sum_{i, j}^{u, c, t}
\lambda_i^{bd}\lambda_j^{sd}
\frac{(m_u^D)^i(m_u^D)^j}{v^2}
 \right|^2
\left[11\left(\frac{1}{m_H^4}+\frac{1}{m_A^4}\right)+\frac{2}{m_H^2 m_A^2}\right].
\qquad
\eea
Using the above formula, we verify that
the correction to $b\to ss\bar{d}$ decay in the semi-aligned 2HDM is given by
$\mathrm{Br}^{\mathrm{FCNH}}[b\to ss\bar{d}]=\mathcal{O}(10^{-14}-10^{-15})$ for
$m_H=10\ \mathrm{TeV}$, which is much smaller than
the SM prediction.
\section{Summary}

We have focused on a situation where the left-right symmetric model with the left-right parity is probed only through
 quark flavor changing neutral currents mediated by the heavy neutral scalar particles (flavor changing neutral Higgses (FCNHs))
 arising from the $SU(2)_L\times SU(2)_R$ bi-fundamental scalar.
We have extracted the bi-fundamental scalar part of the left-right model, named it ``semi-aligned two Higgs doublet model",
 and investigated its phenomenology by focusing on various $\Delta F=2$ processes.

First, we have derived the current lower bound on the mass of FCNH,
 by calculating $\Delta M_{B_d},~\Delta M_{B_s},~\mathrm{Re}\ \epsilon$ and $\sin2\beta_{\mathrm{eff}}$
 with the inclusion of the FCNH exchange contribution,
 and fitting them and other observables with the CKM matrix.
The tension in the fitting, represented by p-values, has given the bounds on the FCNH mass, which read:
$m_H>75-84\ \mathrm{TeV}$ for $3\sigma$ and $m_H>56-64\ \mathrm{TeV}$ for $5\sigma$.

Secondly, we have made a prediction for $(\mathrm{Re}\ \epsilon, \Delta M_{B_d}, \Delta M_{B_s})$ in terms of only one free parameter, i.e.
 the FCNH mass,
 under the assumption that uncertainties in the measurement of $ |V_{ub}|$ and $\gamma$ are reduced from the current estimates,
 allowing us to determine the CKM matrix without being affected by the FCNH exchange.
We have revealed that if the precision of $|V_{ub}|$ and $\gamma$ is improved and the SM prediction disagrees with the experimental data in such a way that the inclusion of FCNH contributions (for some unique value of the FCNH mass) is mandatory to fit the data,
it hints at the semi-aligned two Higgs doublet model and the left-right model.

\section*{Acknowledgement}
We would like to thank Akimasa Ishikawa, who
commented on future experimental precisions of
CKM matrix elements and $\gamma$.
This work is partially supported by Scientific Grants by the Ministry of Education, Culture, Sports, Science and Technology of Japan (Nos. 24540272, 26247038, 15H01037, 16H00871, and 16H02189).
\appendix
\section*{Appendix: Left-right symmetric model with left-right parity}

We present the left-right symmetric model~\cite{lr} with the left-right parity.
The gauge symmetry is $SU(3)_C\times SU(2)_L \times SU(2)_R\times U(1)_{B-L}$,
and the full field content is shown in Table~\ref{fieldcontent}.
The model is invariant under the left-right parity transformation, in which $SU(2)_L$ and $SU(2)_R$ gauge groups are interchanged
 and the fields transform as
\begin{align} 
\Phi \leftrightarrow \Phi^\dagger, \ \ \ \Delta_L \leftrightarrow \Delta_R, \ \ \ q_L \leftrightarrow q_R, \ \ \ \ell_L \leftrightarrow \ell_R.
\end{align}
\begin{table}[h]
\caption{Full field content of the left-right symmetric model. $i=1,2,3$ is the flavor index.}
\begin{center}
\begin{tabular}{|c|c|c|c|c|c|} \hline
Field    & Lorentz $SO(1,3)$ & $SU(3)_C$ & $SU(2)_L\times SU(2)_R$ & $U(1)_{B-L}$ \\ \hline
$q_L^i$& ({\bf 2},~{\bf 1})   &{\bf 3}          & ({\bf 2},~{\bf 1})                 &  1/3  \\
$q_R^i$& ({\bf 1},~{\bf 2})   &{\bf 3}          & ({\bf 1},~{\bf 2})                 &  1/3  \\
$\ell_L^i$& ({\bf 2},~{\bf 1})   &{\bf 1}          & ({\bf 2},~{\bf 1})                 &  $-1$  \\
$\ell_R^i$& ({\bf 1},~{\bf 2})   &{\bf 1}          & ({\bf 1},~{\bf 2})                 &  $-1$  \\ \hline
$\Phi$& {\bf 1}                     & {\bf 1}          & ({\bf 2},~{\bf 2})                 &  0  \\
$\Delta_L$ & {\bf 1}           & {\bf 1}          & ({\bf 3},~{\bf 1})                 &  $2$ \\ 
$\Delta_R$ & {\bf 1}           & {\bf 1}          & ({\bf 1},~{\bf 3})                 &  $2$ \\ \hline
\end{tabular}
\end{center}
\label{fieldcontent}
\end{table}
We write the bi-fundamental scalar $\Phi$ and triplet scalars $\Delta_L,\Delta_R$ as $2\times2$ matrices
that transform under a $SU(2)_L\times SU(2)_R$ gauge transformation as
\begin{align} 
\Phi&\to e^{i\tau^a\theta_L^a}\Phi e^{-i\tau^a\theta_R^a}, 
\ \ \ \Delta_L \to e^{i\tau^a\theta_L^a}\Delta_L e^{-i\tau^a\theta_L^a},
\ \ \ \Delta_R \to e^{i\tau^a\theta_R^a}\Delta_R e^{-i\tau^a\theta_R^a},
\\
&\theta_L^a, \, \theta_R^a:{\rm gauge \ parameters}, \ \ \ \tau^a\equiv \sigma^a/2.
\nonumber
\end{align}

The Yukawa couplings and the scalar potential of the left-right symmetric model with the left-right parity are found to be
\begin{align} 
&-{\cal L}=(Y_q)_{ij} \ \bar{q}_L^i \Phi q_R^j + (\tilde{Y}_q)_{ij} \ \bar{q}_L^i \tilde{\Phi} q_R^j
+(Y_\ell)_{ij}\ \bar{\ell}_L^i \Phi \ell_R^j + (\tilde{Y}_\ell)_{ij}\ \bar{\ell}_L^i \tilde{\Phi} \ell_R^j + {\rm H.c.}
\label{yukawa}\\
&+(Y_M)_{ij} \ \left( \ \ell_L^{i\,T} \epsilon\Delta_L \ell_L^j+\ell_R^{i\,T} \epsilon\Delta_R \ell_R^j \ \right) + {\rm H.c.}
\label{majoranayukawa}\\
&+\mu_1^2 \ {\rm tr}\left[\Phi^\dagger\Phi\right]
+\mu_2^2 \ {\rm tr}\left[\Phi^\dagger\tilde{\Phi}+\Phi\tilde{\Phi}^\dagger\right]
+\mu_3^2 \ {\rm tr}\left[\Delta_L^\dagger\Delta_L+\Delta_R^\dagger\Delta_R\right]
\nonumber \\
&+\lambda_1 \ {\rm tr}\left[\Phi^\dagger\Phi\right]^2
+\lambda_2 \ \left({\rm tr}\left[\Phi^\dagger\tilde{\Phi}\right]^2+{\rm tr}\left[\Phi\tilde{\Phi}^\dagger\right]^2\right)
+\lambda_3 \ {\rm tr}\left[\Phi^\dagger\tilde{\Phi}\right]{\rm tr}\left[\Phi\tilde{\Phi}^\dagger\right]
\nonumber \\
&+\lambda_4 \ {\rm tr}\left[\Phi^\dagger\Phi\right]{\rm tr}\left[\Phi^\dagger\tilde{\Phi}+\Phi\tilde{\Phi}^\dagger\right]
\nonumber \\
&+\rho_1 \ \left({\rm tr}\left[\Delta_L^\dagger\Delta_L\right]^2+{\rm tr}\left[\Delta_R^\dagger\Delta_R\right]^2\right)
+\rho_2 \ \left({\rm tr}\left[\Delta_L\Delta_L\right]{\rm tr}\left[\Delta_L^\dagger\Delta_L^\dagger\right]
+{\rm tr}\left[\Delta_R\Delta_R\right]{\rm tr}\left[\Delta_R^\dagger\Delta_R^\dagger\right]\right)
\nonumber \\
&+\rho_3 \ {\rm tr}\left[\Delta_L^\dagger\Delta_L\right]{\rm tr}\left[\Delta_R^\dagger\Delta_R\right]
+\rho_4 \ \left({\rm tr}\left[\Delta_L\Delta_L\right]{\rm tr}\left[\Delta_R^\dagger\Delta_R^\dagger\right]
+{\rm tr}\left[\Delta_R\Delta_R\right]{\rm tr}\left[\Delta_L^\dagger\Delta_L^\dagger\right]\right)
\nonumber \\
&+\alpha_1 \ {\rm tr}\left[\Phi^\dagger\Phi\right]{\rm tr}\left[\Delta_L^\dagger\Delta_L+\Delta_R^\dagger\Delta_R\right]
\nonumber \\
&+\alpha_{2R} \ 
\left({\rm tr}\left[\Phi^\dagger\tilde{\Phi}\right]{\rm tr}\left[\Delta_L^\dagger\Delta_L\right]
+{\rm tr}\left[\Phi\tilde{\Phi}^\dagger\right]{\rm tr}\left[\Delta_R^\dagger\Delta_R\right]
+{\rm tr}\left[\Phi\tilde{\Phi}^\dagger\right]{\rm tr}\left[\Delta_L^\dagger\Delta_L\right]
+{\rm tr}\left[\Phi^\dagger\tilde{\Phi}\right]{\rm tr}\left[\Delta_R^\dagger\Delta_R\right]\right)
\nonumber \\
&+i \, \alpha_{2I} \ 
\left({\rm tr}\left[\Phi^\dagger\tilde{\Phi}\right]{\rm tr}\left[\Delta_L^\dagger\Delta_L\right]
+{\rm tr}\left[\Phi\tilde{\Phi}^\dagger\right]{\rm tr}\left[\Delta_R^\dagger\Delta_R\right]
-{\rm tr}\left[\Phi\tilde{\Phi}^\dagger\right]{\rm tr}\left[\Delta_L^\dagger\Delta_L\right]
-{\rm tr}\left[\Phi^\dagger\tilde{\Phi}\right]{\rm tr}\left[\Delta_R^\dagger\Delta_R\right]\right)
\nonumber \\
&+\alpha_3 \ {\rm tr}\left[\Phi\Phi^\dagger\Delta_L^\dagger\Delta_L+\Phi^\dagger\Phi\Delta_R^\dagger\Delta_R\right]
\nonumber \\
&+\beta_1 \ {\rm}\left[\Phi\Delta_R\Phi^\dagger\Delta_L^\dagger+\Phi^\dagger\Delta_L\Phi\Delta_R^\dagger\right]
+\beta_2 \ {\rm}\left[\tilde{\Phi}\Delta_R\Phi^\dagger\Delta_L^\dagger+\tilde{\Phi}^\dagger\Delta_L\Phi\Delta_R^\dagger\right]
+\beta_3 \ {\rm}\left[\Phi\Delta_R\tilde{\Phi}^\dagger\Delta_L^\dagger+\Phi^\dagger\Delta_L\tilde{\Phi}\Delta_R^\dagger\right]
\\
&{\rm with} \ \tilde{\Phi}\equiv i\sigma_2 \, \Phi^* \, i\sigma_2,
\nonumber
\end{align}
 where $Y_q,\tilde{Y}_q,Y_\ell,\tilde{Y}_\ell$ are Hermitian matrices, and $Y_M$ is a complex-valued symmetric matrix.
The mass terms $\mu_1^2,\mu_2^2,\mu_3^2$ and the coupling constants 
$\lambda_1,\lambda_2,\lambda_3,\lambda_4,\rho_1,\rho_2,\rho_3,\rho_4,\alpha_1,\alpha_2,\tilde{\alpha}_2,\alpha_3,\beta_1,\beta_2,\beta_3$ are all real.

Through a $SU(2)_R\times U(1)_{B-L}$ symmetry transformation, one can set the VEV of $\Delta_R$ in the following form:
\begin{align} 
\langle \Delta_R \rangle&= \frac{1}{\sqrt{2}}\begin{pmatrix} 
      0 & 0 \\
      v_R & 0 \\
   \end{pmatrix}, \ \ \ \ \ v_R>0.
\end{align}
Through a subsequent $SU(2)_L$ and $\sigma_3$ part of $SU(2)_R$ symmetry transformation, one can set
\begin{align}
\langle \Phi \rangle=   \begin{pmatrix} 
      v_1 & 0 \\
      0 & -e^{i \, \alpha}v_2 \\
   \end{pmatrix}, \ \ \ \ \ v_1>0, \ v_2>0,
\end{align}
 which gives rise to a phase for the $\Delta_R$ VEV, but this can be negated by a $U(1)_{B-L}$ symmetry transformation.


\end{document}